\documentclass[trackchanges, twocolumn]{aastex701}
\usepackage{booktabs}

\newcommand{\BBS}{Building Blocks System}
\newcommand{\Ha}{H$\alpha$}
\newcommand{\Hb}{H$\beta$}
\newcommand{\Hg}{H$\gamma$}
\newcommand{\OIII}{[O{\sc iii}]}

\newcommand{\COMMENT}[1]{\textcolor{purple}{\tt #1\\}}

\defcitealias{Sarrouh2026}{Sarrouh \& Asada et al. (2026)}

\begin{document}

\title{Early steps in the hierarchical assembly of a Milky Way-mass galaxy 1 Gyr after the Big Bang}

\author[0000-0003-3983-5438]{Yoshihisa Asada}
\affiliation{Dunlap Institute for Astronomy and Astrophysics, 50 St. George Street, Toronto, ON, M5S 3H4, Canada}
\affiliation{Department of Astronomy and Physics and Institute for Computational Astrophysics, Saint Mary's University, 923 Robie Street, Halifax, NS, B3H 3C3, Canada}
\affiliation{Waseda Research Institute for Science and Engineering, Faculty of Science and Engineering, Waseda University, 3-4-1 Okubo, Shinjuku, 169-8555, Tokyo, Japan}
\email[show]{yoshi.asada@utoronto.ca}

\author[0000-0002-7712-7857]{Marcin Sawicki}
\email{marcin.sawicki@smu.ca}
\affiliation{Department of Astronomy and Physics and Institute for Computational Astrophysics, Saint Mary's University, 923 Robie Street, Halifax, NS, B3H 3C3, Canada}

\author[0000-0001-8325-1742]{Guillaume Desprez}
\email{guillaume.desprez@protonmail.com}
\affiliation{Department of Astronomy and Physics and Institute for Computational Astrophysics, Saint Mary's University, 923 Robie Street, Halifax, NS, B3H 3C3, Canada}.
\affiliation{Kapteyn Astronomical Institute, University of Groningen, P.O. Box 800, 9700AV, Groningen, The Netherlands}

\author[0009-0000-2101-1938]{Jon Jude\v{z}}
\affiliation{Faculty of Mathematics and Physics, University of Ljubljana, Jadranska ulica 19, Ljubljana, SI-1000, Slovenia}
\email{jon.judez@fmf.uni-lj.si}

\author[0000-0001-5984-0395]{Maru\v{s}a Brada\v{c}}
\affiliation{Faculty of Mathematics and Physics, University of Ljubljana, Jadranska ulica 19, Ljubljana, SI-1000, Slovenia}
\affiliation{Department of Physics and Astronomy, University of California Davis, 1 Shields Avenue, Davis, 95616, CA, USA}
\email{marusa.bradac@fmf.uni-lj.si}

\author[0000-0001-9298-3523]{Kartheik Iyer}
\affiliation{Columbia Astrophysics Lab, Columbia University, 550 W 120th St, New York, 10010, NY, USA}
\email{kgi2103@columbia.edu}

\author[0000-0003-3243-9969]{Nicholas S. Martis}
\affiliation{Faculty of Mathematics and Physics, University of Ljubljana, Jadranska ulica 19, Ljubljana, SI-1000, Slovenia}
\email{nicholas.martis@fmf.uni-lj.si}

\author[0000-0002-9330-9108]{Adam Muzzin}
\affiliation{Department of Physics and Astronomy, York University, 4700 Keele St., Toronto, ON, M3J 1P3, Canada}
\email{muzzin@yorku.ca}

\author{Ga\"el Noirot}
\affiliation{Space Telescope Science Institute, 3700 San Martin Drive, Baltimore, 21218, MD, USA}
\affiliation{Department of Astronomy and Physics and Institute for Computational Astrophysics, Saint Mary's University, 923 Robie Street, Halifax, NS, B3H 3C3, Canada}
\email{gnoirot@stsci.edu}

\author[0000-0001-8830-2166]{Ghassan T. E. Sarrouh}
\affiliation{Department of Physics and Astronomy, York University, 4700 Keele St., Toronto, ON, M3J 1P3, Canada}
\email{gsarrouh@yorku.ca}

\author[0000-0002-4201-7367]{Chris J. Willott}
\affiliation{Herzberg Astronomy \& Astrophysics Research Centre, National Research Council of Canada, 5071 West Saanich Road, Victoria, BC, V9E 2E7, Canada}
\email{chris.willott@nrc.ca}

\author[0000-0002-0243-6575]{Jacqueline Antwi-Danso}
\affiliation{David A. Dunlap Department of Astronomy and Astrophysics, University of Toronto, 50 St. George Street, Toronto, ON, M5S 3H4, Canada}
\email{j.antwidanso@utoronto.ca}

\author[0000-0001-7201-5066]{Seiji Fujimoto}
\affiliation{David A. Dunlap Department of Astronomy and Astrophysics, University of Toronto, 50 St. George Street, Toronto, ON, M5S 3H4, Canada}
\affiliation{Dunlap Institute for Astronomy and Astrophysics, 50 St. George Street, Toronto, ON, M5S 3H4, Canada}
\email{seiji.fujimoto@utoronto.ca}

\author[0000-0001-9002-3502]{Danilo Marchesini}
\affiliation{Department of Physics \& Astronomy, Tufts University, 574 Boston Avenue, Medford, 02155, MA, USA}
\email{danilo.marchesini@tufts.edu}

\author[0000-0002-5694-6124]{Vladan Markov}
\affiliation{Faculty of Mathematics and Physics, University of Ljubljana, Jadranska ulica 19, Ljubljana, SI-1000, Slovenia}
\email{vladan.markov@fmf.uni-lj.si}

\author[0000-0002-6265-2675]{Luke Robbins}
\affiliation{Department of Physics \& Astronomy, Tufts University, 574 Boston Avenue, Medford, 02155, MA, USA}
\email{andrew.robbins@tufts.edu}

\author[0000-0003-0780-9526]{Visal Sok}
\affiliation{Department of Astrophysical and Planetary Sciences, University of Colorado, 2000 Colorado Ave, Boulder, CO 80309, USA}
\email{visal.sok@colorado.edu}

\begin{abstract}

We report JWST observations of a $z_{\rm spec}=5.196$ compact group of three strongly lensed low-mass galaxies ($M_\star \sim 10^{6-7} M_\odot$ each) whose small  line-of-sight velocity offsets (from $-120\pm100$ to  $+160\pm110$ km\! s$^{-1}$) and projected separations ($\sim2$ kpc) suggest that they are undergoing merging. On the basis of abundance-matching arguments, the trio appear to be destined to evolve into a Milky Way-mass galaxy by the present day. Our spectrophotometric analysis suggests that the stellar mass growth of this system is not simply just due to the merging of its components; rather, it is being dramatically enhanced by intense bursts of star formation. Presumably initiated by tidally-induced gas inflows, these  star-forming bursts boost the mass growth $2.6\pm0.5$ times that expected in straightforward merging of the existing stellar masses. 
These spectroscopic observations thus not only demonstrate that hierarchical assembly remains a viable formation channel in the early assembly phases of present-day massive galaxy, but also that interaction-induced bursts of star formation are a major accelerator of this assembly process.

\end{abstract}

\keywords{\uat{Emission line galaxies}{459} --- \uat{Galaxy evolution}{594} --- \uat{Galaxy interactions}{600} -- \uat{Gravitational lensing}{670} -- \uat{High-redshift galaxies}{734}}


\section{Introduction} 

In the hierarchical picture of structure formation \citep[e.g.,][]{White1991,Lacey1993}, galaxies like the Milky Way (MW)  are expected to have grown through successive mergers. Hierarchical galaxy growth has been observationally confirmed in the low-$z$ universe \citep[e.g.,][]{Patton2002}, and it is now possible to investigate this assembly process in its early stages by observing low-mass high-$z$ galaxies with the James Webb Space Telescope \citep[JWST;][]{Gardner2023}.

Progenitors of MW-like galaxies in the first billion years of cosmic history ($z\gtrsim5$) are expected to be very faint and have low stellar masses \citep[$M_\star \lesssim10^8\ M_\odot$; see e.g.,][]{Mowla2024Natur}.
Amplification by strong gravitational lensing has therefore been often used to study the detailed physical properties of such faint high-$z$ systems \citep[e.g.,][]{Adamo2024Natur,Mowla2024Natur,Fujimoto2024arxiv}.
In particular, spectroscopic observations of these lensed galaxies with the JWST/NIRSpec instrument \citep{Jakobsen2022} has been used to reveal their interstellar medium (ISM) properties and star formation histories, shedding light on the physics behind the early galaxy evolution \citep[e.g.,][]{Chemerynska2024ApJ,Atek2024Natur,Asada2026arXiv}. Notably, JWST observations of several high-$z$ lensed galaxies have reported clumpy structure embedded within a diffuse disk-like component \citep[e.g.,][]{Adamo2024Natur,Mowla2024Natur,Fujimoto2024arxiv,Bradac2025ApJ,Nakane2025arXiv}.
These studies suggest that \textit{in-situ} massive star cluster formation
may be the main channel of stellar mass growth in these faint high-$z$ galaxies, which stands at odds with the classical hierarchical assembly paradigm.

In contrast to star-forming clump scenario, galaxy-galaxy mergers represent a more direct formation channel for hierarchical galaxy assembly. Several statistical studies have found the fraction of galaxy pairs -- possible merger progenitors -- at high-$z$ ($z\gtrsim5$) to be $\sim20-40$\% \citep{Ventou2017, Duncan2019, Asada2024MNRAS, Stephenson2025,Duan2025,Puskas2025}. These studies suggest that star formation may be impacted by galaxy-galaxy interactions, particularly at small projected separations \citep[e.g.,][]{Asada2024MNRAS, Puskas2025b, Duan2026, Omori2026arXiv}. While such  statistical pair studies suggest that galaxy-galaxy interactions may be important at high-$z$, little {\it direct} evidence of hierarchical assembly in low-mass galaxies has been reported to date and the details of the role of interactions and/or mergers on the physics of early stellar mass assembly are not yet well explored.  

\begin{figure*}[tb]
\centering
\includegraphics[width=0.9\textwidth]{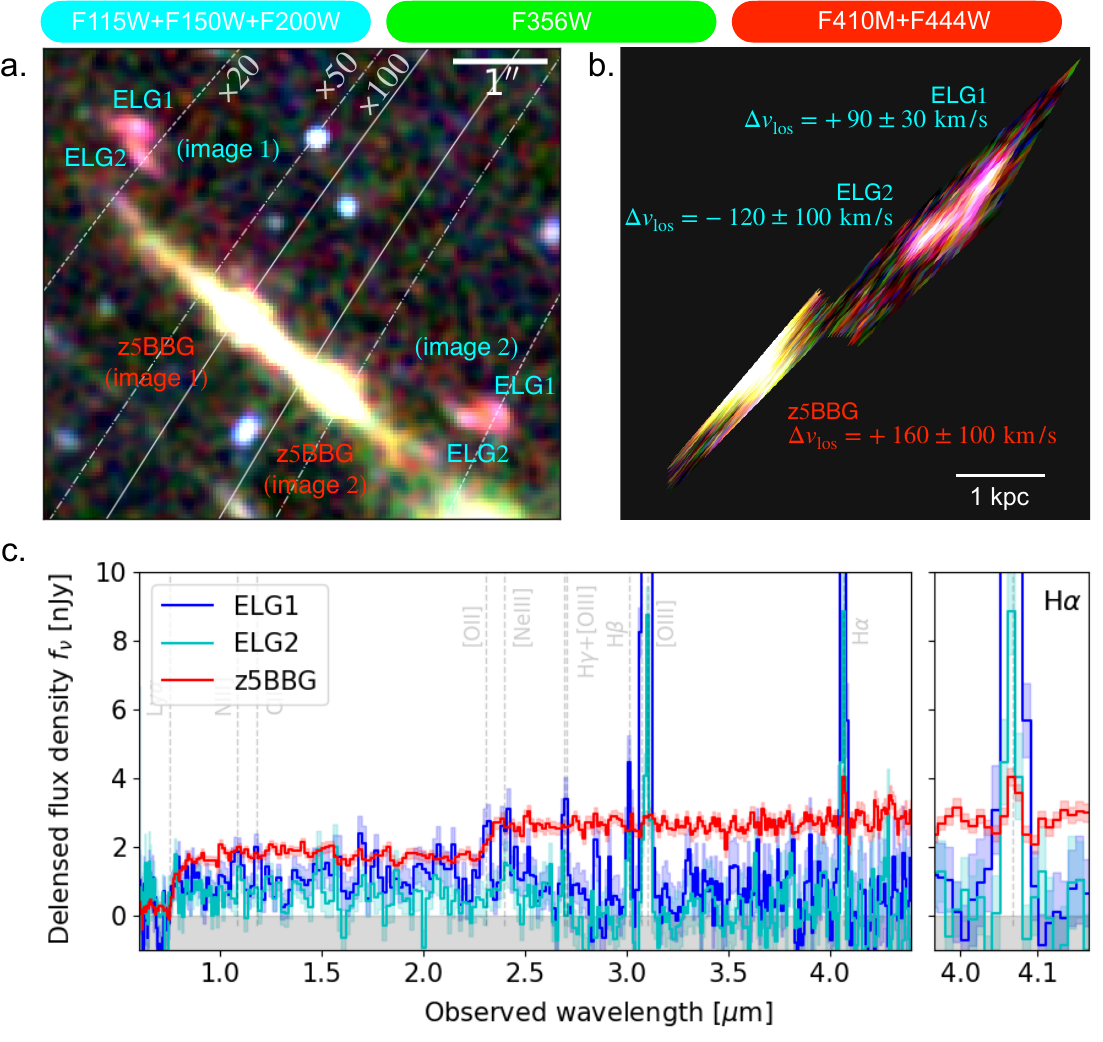}
\caption{The \BBS, a low-mass galaxy group at $z_{\rm spec}=5.196$ behind the lensing cluster MACS~0417.
{\bf a.} RGB image of the \BBS. Contours show the iso-magnification curves ($\mu=20$, $50$, and $100$) of the gravitational lensing. The three galaxies (ELG1, ELG2, and z5BBG) are doubly imaged due to the lensing. The purple colors of ELG1 and ELG2 are due to the presence of strong H$\alpha$ in the F410M filter; z5BBG has only very weak H$\alpha$ emission, making it appear whiteish in this color composite image. 
{\bf b.} Source-plane reconstructed RGB image of the system. The diagonal elongation of the images is the artifact caused by the lensing shear of the telescope’s point spread function (PSF) when projected from the observed plane to the source plane.
{\bf c.} The NIRSpec/PRISM 1D spectra of the three galaxies. Flux densities are corrected for gravitational lens magnification. The \Ha\ lines in all three galaxies are at the same wavelength, confirming the three sources are at the same redshift. However, the spectra show obvious differences in their spectral types.
}\label{fig1}
\end{figure*}

One of the best laboratories to investigate in detail the high-$z$ mass assembly via high-$z$ galaxy-galaxy interactions lies behind the cluster MACS J0417.5-1154 (hereafter MACS0417) and consists of three gravitationally-lensed low-mass galaxies shown in Figure~\ref{fig1}. As we show in this paper, the three galaxies lie at a common redshift of $z_{\mathrm{spec}}=5.20$ and are likely to be interacting. The trio were originally thought to be two distinct systems at different redshifts.  Two of the three, named MACS0417-ELG1 and MACS0417-ELG2 (ELG1 and ELG2 hereafter) and both with $M_\star<10^7\ M_\odot$, were first reported by \citet{Asada2023MNRAS} at $z_{\rm phot}\sim5.1$ based on NIRCam imaging that showed a strong emission-line excess in the F410M medium band filter. ELG1 and ELG2 have stellar masses $M_\star<10^7\ M_\odot$, and both show photometric SEDs consistent with strong emission lines of [O{\sc iii}]+H$\beta$ and H$\alpha$ indicative of intense, recently-started bursts of star formation \citep[][]{Asada2023MNRAS}. On the other hand, \citet{Strait2023ApJ} reported another lensed galaxy behind MACS0417 but with a prominent Balmer break and a very weak H$\alpha$ emission line in its NIRSpec/Prism spectrum.  Named MACS0417-z5BBG (z5BBG hereafter), specro-photometric analysis showed it to be a low-mass ($M_\star\sim10^7\ M_\odot$) galaxy caught soon after the quenching of its star formation. 

In this paper, we present NIRSpec/Prism observations of ELG1 and ELG2 which prove that this star-bursting pair is at the same redshift as the recently quenched galaxy z5BBG. The trio thus constitute a compact $z_{\rm spec}=5.20$, interacting galaxy group which we dub the  ``Building Blocks System'' (BBS). The three members of this system are all highly magnified ($\mu>20$) low-mass ($M_\star\lesssim10^7\ M_\odot$) galaxies, making them an ideal laboratory to study the mass assembly history in what appears to be a system undergoing hierarchical assembly $\sim$1~Gyr after the Big Bang. Together with the previously presented NIRSpec observations of z5BBG, we jointly analyze the spectro-photometric data of the three member galaxies of the BBS group and derive the mass assembly history of this system.

The paper is structured as follows.
We first briefly describe the data and observations used in this work in Section   \ref{sec:data}. We then show the spectro-photometry data analyses in Section \ref{sec:analysis}.
In Section~ \ref{sec:results} we present the key results of this paper, discussing the stellar mass assembly of the BBS and the role of galaxy-galaxy interactions among lowest-mass galaxies.
Throughout this paper we assume the $\Omega_{m,0}=0.3, \Omega_{\Lambda,0}=0.7, H_0=70$ km/s/Mpc cosmological model. Magnitudes are on the AB system, and the Chabrier stellar initial mass function (IMF; \citealt{Chabrier2003}) is adopted.

\section{Data}\label{sec:data}


\subsection{Imaging}

We used JWST/NIRCam and NIRISS imaging observations of the MACS0417 cluster field taken as part of the CANUCS Guaranteed Time Observations program \cite{Willott2022PASP}, as well as HST/ACS and WFC3 imaging observations (HST-GO-16667 PI Brada\v{c}) that we retrieved from the MAST data archive. The overall survey design and data reduction process is detailed in \citetalias{Sarrouh2026}, and we summarize the key elements here. 

NIRCam imaging observations were taken through the F090W, F115W, F150W, F200W, F277W, F356W, F410M, and F444W filters with 6.4 ks exposure time each, while NIRISS direct imaging mode observations in the F115W, F150W, and F200W filters had 2.3 ks exposures each. The images were processed with a combination of the official STScI pipeline and the \texttt{grizli} software \cite{Brammer2019ascl}. The astrometric solution was assigned to each exposure using the Gaia DR3 catalog and the individual exposures in each filter were drizzled onto a common pixel grid with 40 milliarcsec pixels. Bright cluster galaxies were modeled and subtracted using a custom code described by \cite{Martis2024ApJ}. Point spread functions (PSFs) for each filter were generated from isolated stars in the field, and the drizzled image stacks were convolved with a kernel designed to homogenize the PSFs in all the images to that of the lowest resolution band, F444W.

\subsection{Spectroscopy}

\begin{figure}[t]
\centering
    \includegraphics[width=1\linewidth]
{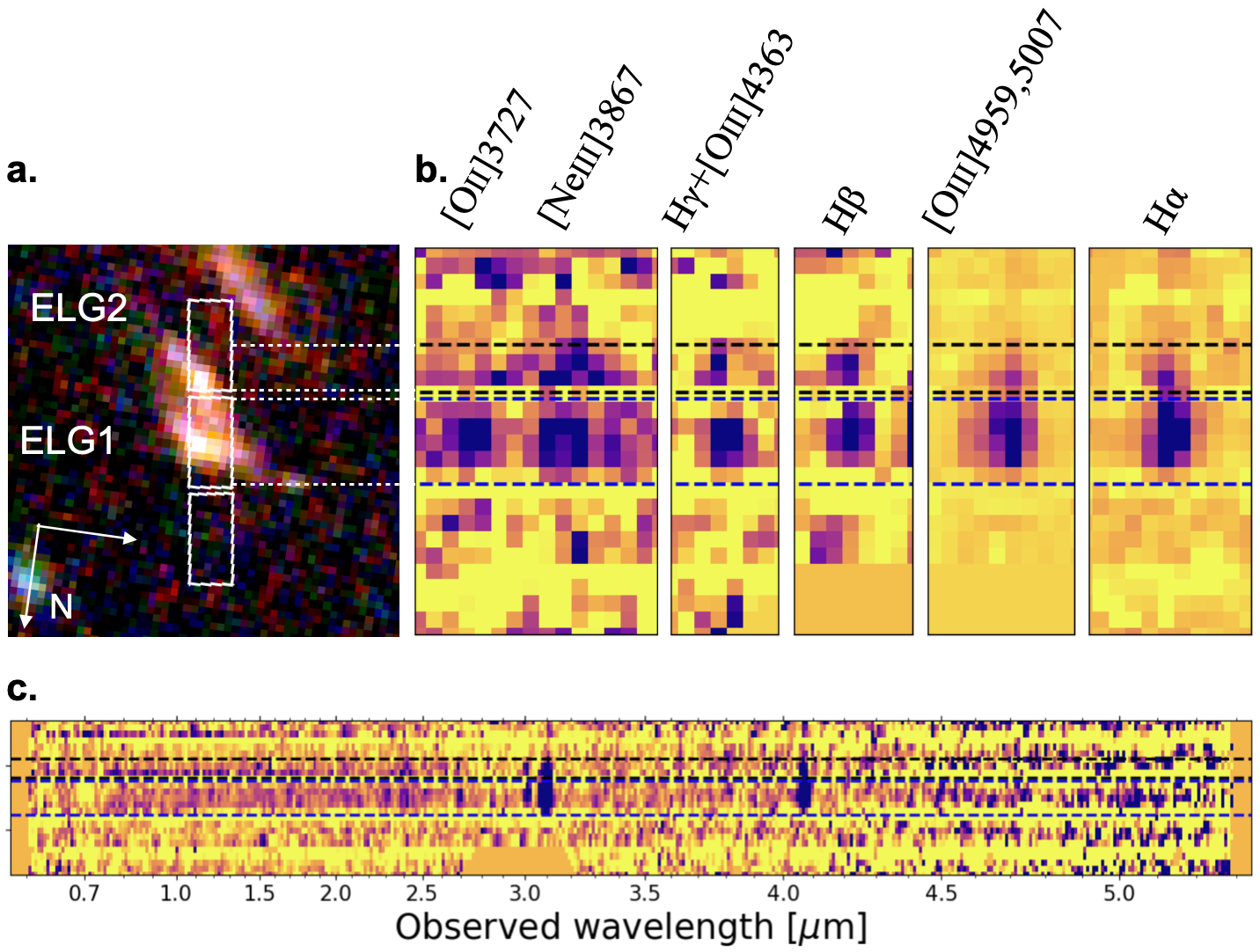}
\caption{The JWST/NIRSpec observation of the ELG1+ELG2 galaxy pair. 
\textbf{a.} The slitlet position of the NIRSpec observation.
The RGB image is rotated by the position angle of the observation (188.1$^\circ$),
so that the x-axis of this image corresponds to the dispersion direction on the NIRSpec detector and tye y-axis correspoinds to the cross-dispersion direction.
The direction of dispersion is from left to right in this coordinate system.
\textbf{b.} The zoom-in 2D spectrum around emission lines. From left to right, 2D spectra around [O{\sc ii}]$\lambda$3727 and [Ne{\sc iii}]$\lambda$3867, H$\gamma$+[O{\sc iii}]$\lambda$4363, H$\beta$, [O{\sc iii}]$\lambda\lambda$4959,5007, and H$\alpha$ are shown.
\textbf{c.} The full 2D spectrum of the ELG1+ELG2 pair. The dashed lines denote the position of ELG1 and ELG2.
}\label{fig:ELG12_2Dspec}
\end{figure}

We used NIRSpec multi-object spectroscopic observations taken with PRISM/CLEAR as a part of the CANUCS program. The spectroscopic observations were performed using the NIRSpec Micro-Shutter Assembly (MSA; \citealt{Ferruit2022}), assigning a 3-shutter-tall slitlet to each target.  Targets were selected to cover various science cases of the CANUCS program based on the photometric catalog from the CANUCS Cycle 1 NIRISS and NIRCam observations supplemented by archival HST optical observations.
In this work, we focus on the group of three galaxies ELG1, ELG2, and z5BBG (the \BBS). 



Our NIRSpec spectroscopic data were processed with a combination of standard STScI pipeline (software version 1.8.4 and (jwst\_1030.pmap) and the msaexp package \cite{Brammer2022zndo}. The level 1 data processing using the STScI standard pipeline to obtain the rate fits file from the raw data was done in the same way as that in \cite{Desprez2024MNRAS}. We then used msaexp to do level 2 processing where we applied the $1/f$ correction, identified the snowball artifacts on the rate fits file, removed bias from each exposure, performed standard wavelength calibration, flat-fielding, path-loss correction, and photometric calibration. We then drizzle-combined the three nodded exposures to obtain a full 2D spectrum for each source before performing background subtraction using a custom procedure described in Appendix~\ref{appendix:spec}. The resulting 2D spectrum of ELG1 and ELG2 is shown in Figure~\ref{fig:ELG12_2Dspec}.  The 1D spectra of these two objects, toghether with that of z5BBG from \cite{Strait2023ApJ}, are in Figure \ref{fig1}c and show two contrasting sets of characteristics representing both star-forming and quenching objects. 
 
\section{Analysis}\label{sec:analysis}

\subsection{Ionized ISM properties from NIRSpec spectroscopy}
\label{sec:spectral-analysis}

\subsubsection{The galaxy pair ELG1+ELG2}

\begin{figure*}[t]
\centering
\includegraphics[width=0.75\textwidth]{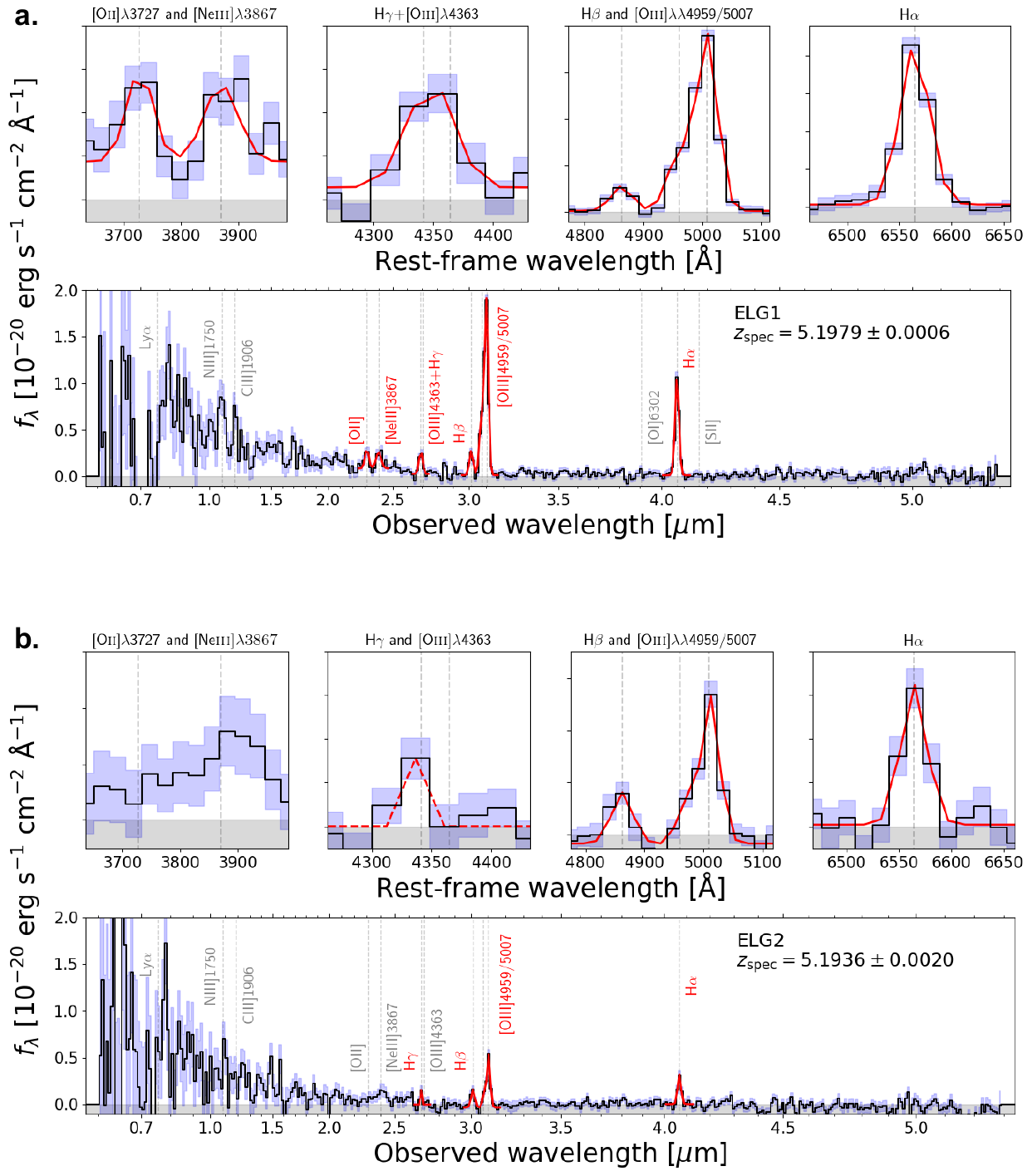}
\caption{The extracted 1D spectra of ELG1 (panel a) and ELG2 (panel b) galaxy. The full 1D spectrum is shown in the bottom row, and the zoom-in spectra around key emission lines are shown in top row sub-panels. The black solid lines present the observed 1D spectrum with 1$\sigma$ uncertainties shaded in blue, and the red lines present the best-fit Gaussian profiles of detected emission lines. Gray vertical dashed lines mark the wavelengths of strong emission lines.}
\label{fig:ELG12_1Dspec}
\end{figure*}

\begin{figure*}[t]
\centering
\includegraphics[width=0.95\textwidth]{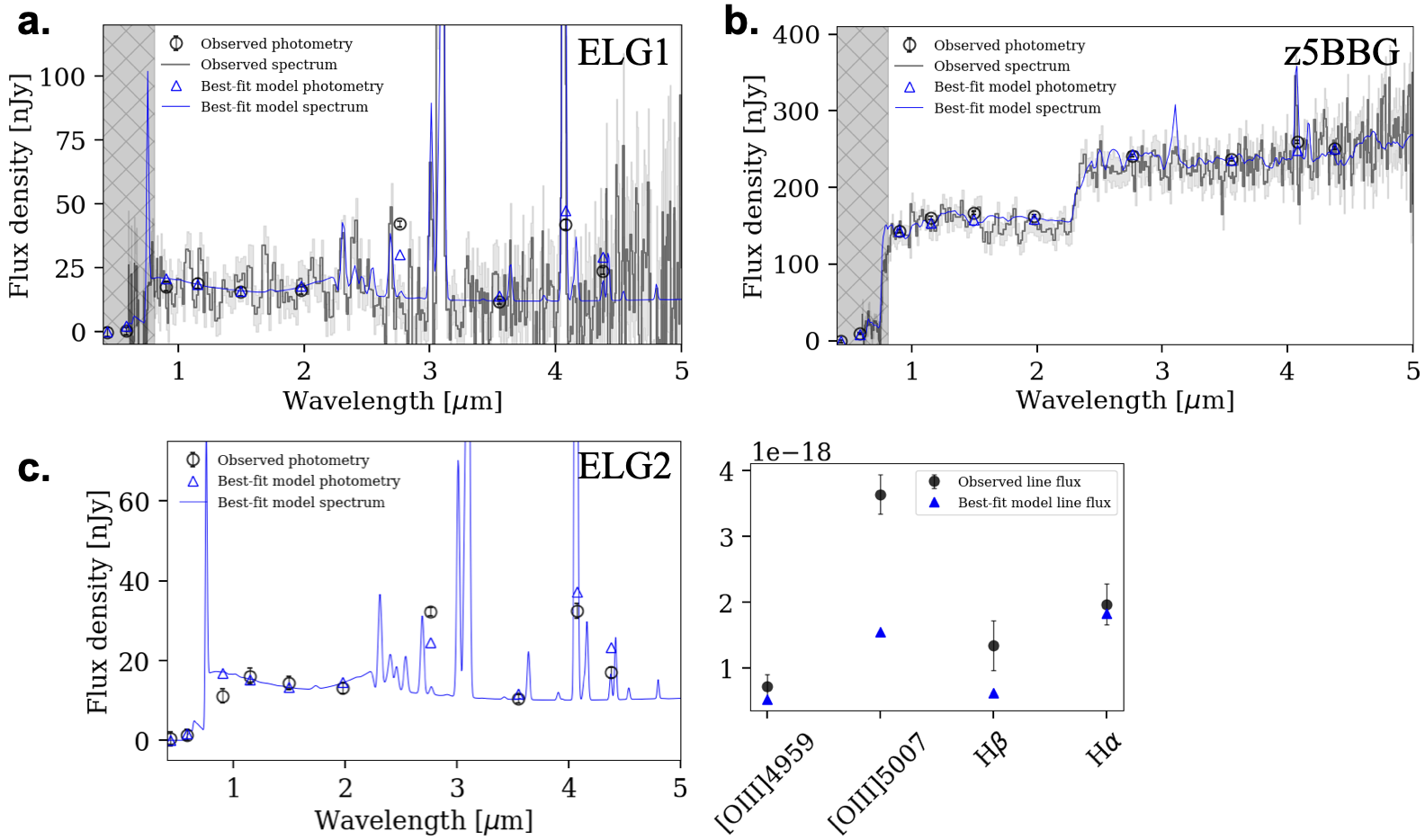}
\caption{\texttt{Prospector} spectrophotometry fitting of ELG1, ELG2, and z5BBG. Black circles with error bars show the observed photometry and its uncertainty, and the gray solid lines with shaded areas are the NIRSpec/PRISM spectra. The blue solid curve is the best-fit model spectrum, and its synthesized model photometry is shown with blue triangles. When the emission line fluxes are used in the fit instead of the full spectrum (ELG2; panel c), the fits to line fluxes are also presented in the subpanel. The inferred SFHs from these fits are shown in Figure \ref{fig:SFHs}. 
}
\label{fig:Prospector_SEDs}
\end{figure*}


Figure \ref{fig:ELG12_2Dspec}c shows the full 2D spectrum of the ELG1+ELG2 pair, and Figure \ref{fig:ELG12_1Dspec} shows the extracted 1D spectra for ELG1 and ELG2 individually.
The 2D spectrum (Fig.~\ref{fig:ELG12_2Dspec}c) shows redshifted emission lines of [O{\sc ii}]$\lambda$3727, [Ne{\sc iii}]$\lambda$3867, H$\gamma$+[O{\sc iii}]$\lambda$4363, H$\beta$, [O{\sc iii}]$\lambda\lambda$4959/5007, and H$\alpha$. 
Figure \ref{fig:ELG12_2Dspec}b presents the zoomed-in 2D spectra around these lines.
The 2D spectrum shows an obvious velocity offset between the two galaxies in all the Balmer line profiles (H$\alpha$, H$\beta$, and H$\gamma$). The spectroscopic redshifts measured with the H$\alpha$ lines are $z_{\rm spec} = 5.1979 \pm 0.0006$ for ELG1 and $z_{\rm spec} = 5.1936 \pm 0.0020$ for ELG2, and the velocity offset between ELG1 and ELG2 is $\Delta v = 203\pm121\ {\rm km\ s^{-1}}$. 

Although the velocity offsets in the Balmer lines are all consistent with each other and seem to be robust, the offset cannot be seen in the [O{\sc iii}]4959,5007 doublet line (Figure \ref{fig:ELG12_2Dspec}b). The difference between Balmer lines and the [O{\sc iii}] doublet suggests that the O$^{++}$ gas velocity profile differs from that of the H$^{+}$ gas.
\cite{Asada2023MNRAS} exploited the NIRCam observation of this system to generate photometry-based emission line maps and argued that the H$\alpha$ and [O{\sc iii}]+H$\beta$ distributions could be different. The velocity structure difference observed in the NIRSpec spectrum is consistent with their  NIRCam result, and supports the idea that the ELG1+ELG2 system has an internal (velocity) structure with different ISM properties, possibly arising from the interaction between ELG1 and ELG2. 

\subsubsection{Emission line analysis of the ELG1 spectrum}

\begin{table*}[tb]
\centering
\caption{ISM properties of ELG1. Line fluxes are in units of $10^{-20}\ {\rm erg\ s^{-1}\ {\rm cm}^{-2}}$.}
\label{tab:Emlines}
\begin{tabular}{@{}l r@{\,\ensuremath{\pm}\,}l|lc@{}}
\toprule
Line & \multicolumn{2}{c}{flux} & Property & \\
\midrule
H$\alpha$    & 225.0 & 9.3  & $z_{\rm spec}$ & 5.1979 $\pm$ 0.0006  \\
\lbrack O{\sc iii}\rbrack 5007    & 440.0 & 13.0 & $T_e({\rm O^{++}})$ & $>2.1\times10^4$ K  \\
\lbrack O{\sc iii}\rbrack4959    & 154.6 & 12.7 & 12+$\log({\rm O/H})$ & $<7.63$  \\
H$\beta$    &  61.5 & 10.2 & $E(B-V)$ & 0.21 $\pm$ 0.15  \\
H$\gamma$+[O{\sc iii}\rbrack4363    &  60.0 & 15.8 &  &   \\
\lbrack Ne{\sc iii}\rbrack3867    & 65.2 & 18.0 &  &   \\
\lbrack O{\sc iii}\rbrack3727    & 60.3 & 15.8 &  &   \\
\bottomrule
\end{tabular}
\end{table*}

The extracted 1D spectrum is corrected for the slit-loss by scaling the NIRSpec spectrum to the PSF-homogenized NIRCam fixed aperture photometry. We fixed the redshift to the H$\alpha$ spec-$z$, and fitted Gaussian profiles to the other emission lines to derive their emission line fluxes.

In the ELG1 spectrum (Figure \ref{fig:ELG12_1Dspec}a), we detect [O{\sc ii}]$\lambda$3727, [Ne{\sc iii}]$\lambda$3867, H$\gamma$+[O{\sc iii}]$\lambda$4363, H$\beta$, [O{\sc iii}]$\lambda\lambda$4959/5007, and H$\alpha$.
With  PRISM resolution ($R\sim100$) it is impossible to resolve the H$\gamma$ and [O{\sc iii}]$\lambda$4363 lines at $\sim2.7\ \mu$m. The 1D spectrum shows the peak is at the wavelength between H$\gamma$ and [O{\sc iii}]$\lambda$4363, and the observed spectral profile cannot be well explained by the presence of only one of the two lines.
For the H$\gamma$+[O{\sc iii}]$\lambda$4363 blended line, we thus fitted two Gaussian components corresponding to the two lines, and derived the total flux of the blended line.

We next derived diagnostic emission line ratios based on the line flux measurements, correcting for dust attenuation estimated from the Balmer decrement. The observed H$\alpha$/H$\beta$ ratio is $3.66\pm0.63$, which gives a modest dust attenuation of $E(B-V)=0.21\pm0.15$ assuming the intrinsic ratio of $2.86$ under case B recombination \citep{Osterbrock2006agnagn} and Calzetti dust \citep{Calzetti2000}. We use this measured attenuation to correct the extracted line fluxes.

Next, we  deblend the H$\gamma$ and [O{\sc iii}]4363 lines in order to measure the temperature-sensitive emission line ratio [O{\sc iii}]$_{4959+5007}$/[O{\sc iii}]$_{4363}$.
In spite of the blending, the 1D line profile of the H$\gamma$+[O{\sc iii}]4363 blended line cannot be well explained by the H$\gamma$ line only, and suggests some contribution from the [O{\sc iii}]4363 line component. The blended line flux of H$\gamma$+[O{\sc iii}]4363 is significantly larger than that predicted for H$\gamma$ from H$\beta$ and H$\alpha$: the (dust- and slit-loss-corrected) line ratios of (H$\gamma$+[O{\sc iii}]4363)/H$\beta$ are $1.08 \pm 0.34$, while H$\gamma$/H$\beta$ should be 0.47 intrinsically \citep{Osterbrock2006agnagn}, indicating a non-negligible contribution from the [O{\sc iii}]4363 line. We thus subtracted the predicted H$\gamma$ flux from the blended H$\gamma$+[O{\sc iii}]4363 flux to obtain the [O{\sc iii}]4363 line flux.
Based on the [O{\sc iii}]4363 line flux measurement, we obtained the RO3(=[O{\sc iii}]$_{4959+5007}$/[O{\sc iii}]$_{4363}$) line ratio of $15.4\pm7.2$.

Finally, we measured the oxygen abundance in ELG1's ionized gas with the direct temperature method \citep{Izotov2006}, following the standard prescription used in other high-$z$ JWST galaxy studies \citep{Curti_2023MNRAS,Sanders_2024ApJ,Mowla2024Natur,Markov2025arXiv}. Namely, we compute the electron temperature in the O$^{++}$ emitting region $T_e({\rm O^{++}})$ from the RO3 line ratio assuming an electron density consistently with literature \citep{Isobe2023}, infer the temperature in the O$^{+}$ emitting region $T_e({\rm O^{+}})$ based on the \citet{Campbell1986} relation between $T_e({\rm O^{++}})$ and $T_e({\rm O^{+}})$, and estimate the total oxygen abundance (O/H) from O$^{++}$/H and O$^{+}$/H assuming higher ionizing state oxygen is negligible \citep{Berg2021}.
In the case of the ELG1 spectrum, considering the low S/N of the RO3 ratio and the blending of the H$\gamma$ and [O{\sc iii}]4363 lines, we only report the 3-sigma lower limit of the electron temperature $T_e$ and the 3-sigma upper limit of oxygen abundance. The observed emission line fluxes and the ISM properties derived for ELG1 are presented in Table \ref{tab:Emlines}.

\subsubsection{The ELG2 spectrum}

In the ELG2 spectrum (Figure \ref{fig:ELG12_1Dspec}b), we detected H$\beta$, [O{\sc iii}]$\lambda\lambda$4959/5007, and H$\alpha$, and found a tentative detection of H$\gamma$. The ELG2 galaxy was located at the edge of the shutter and suffered from heavy slit-losses, which makes the observed spectrum faint and noisy. Although the H$\gamma$ line is detected at the expected wavelength in the 1D spectrum and the 2D spectrum seems to show the signal of this line, the S/N of the peak is only $\sim3.3$. In addition, some flux excess around [Ne{\sc iii}]$\lambda$3867 could be seen  in both the 1D and 2D spectra, but the line profile is too noisy and the peak is not well aligned with the expected wavelength. It is thus impossible to derive the key ISM properties ($T_e$ and O/H) for ELG2 due to the lack of [O{\sc ii}]3727 and [O{\sc iii}]4363 line detection.



\subsection{Physical properties from spectrophotometric SED fitting}
\label{sec:SEDfitting}

\begin{table*}[tb]
\centering
\caption{Physical properties of the \BBS\ components.}\label{tab:Phys_params}%
\begin{tabular}{@{}lccc@{}}
\toprule
Properties & ELG1  & ELG2 & z5BBG \\
\midrule
$z_{\rm spec}$    & $5.1979 \pm 0.0006$ & $5.1936 \pm 0.0020$ & $5.1993 \pm 0.0023$   \rule{0pt}{1.2em} \\
$\mu_{\rm best}$    & $17.5^{+0.6}_{-0.6}$ & $19.7^{+0.6}_{-0.9}$ & $104.7^{+24.3}_{-16.6}$   \rule{0pt}{1.2em} \\
$\log(M_\star/M_\odot)$    & $6.46^{+0.03}_{-0.02}$ & $6.03^{+0.08}_{-0.05}$ & $7.54^{+0.01}_{-0.01}$   \rule{0pt}{1.2em} \\
$\log({\rm sSFR/Gyr^{-1}})$    & 1.99$^{+0.04}_{-0.03}$ & 1.95$^{+0.19}_{-0.18}$ & -1.94$^{+0.58}_{-1.11}$  \rule{0pt}{1.2em} \\
\bottomrule
\end{tabular}
\end{table*}

We derived the physical parameters of ELG1, ELG2, and z5BBG by performing spectro-photometric SED fitting to the NIRSpec PRISM spectra together with HST+NIRISS+NIRCam photometry with {\tt Prospector}. We used the Binary Population and Spectral Synthesis code (BPASS; \citealt{Eldridge2017}) as the SSP model spectrum, and utilized a non-parametric SFH (``{\tt continuity\_sfh}") from \cite{Leja2019} with 6 age bins in the fit. The age bins were put at 0-10 Myr, 10-30 Myr, and the other four bins were logarithmically spaced up to $z=20$.
We also included nebular emission in the model spectrum, whose metallicity and ionization parameter are varied as free parameters. The redshift was fixed to the spec-$z$ value. With this setup, the free parameters are the stellar mass, dust attenuation, metallicity, ionization parameter, and five SFR ratios between the adjacent age bins.

We fit the model spectrum to the NIRSpec spectrum and HST+NIRISS+NIRCam photometry simultaneously. We did not use HST/WFC3 IR photometry in the fit since the wavelength coverage overlaps with JWST/NIRCam but the S/Ns are significantly lower than NIRCam. In the spectral fitting, we fit to the full spectrum only when the continuum is clearly detected in the observed spectrum (i.e., ELG1 and z5BBG) since a noisy continuum is not informative in the fit. When the continuum is not clearly detected (ELG2), we fit to the observed emission line fluxes instead of the full observed spectrum. We also masked out $\lambda_{\rm rest}<1500$ \AA\ in the full spectrum fitting to avoid the uncertainties in the modeling of the Ly$\alpha$ line escape fraction and damping wing absorption.

\begin{figure*}[tb]
\centering
\includegraphics[width=0.9\textwidth]{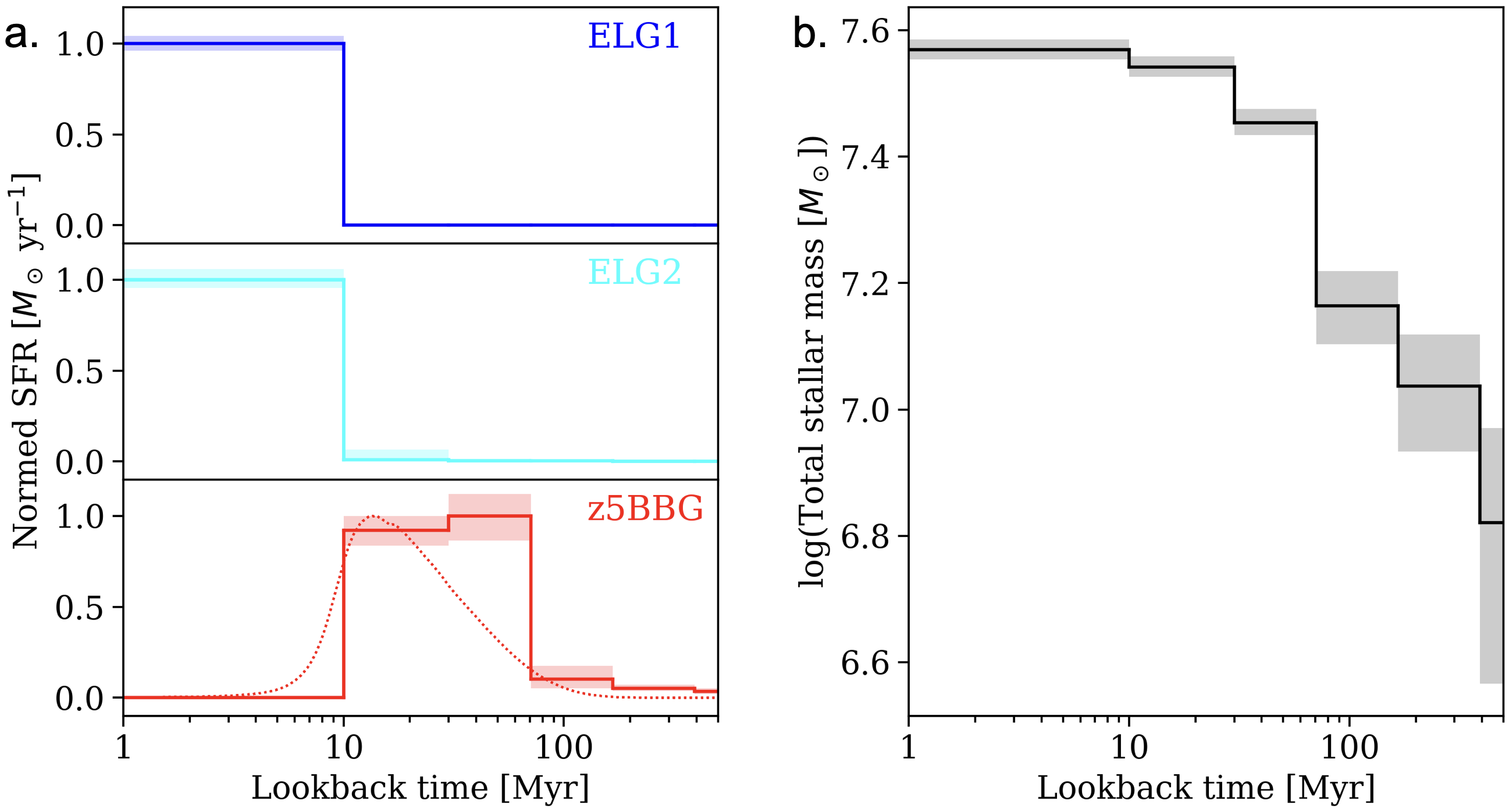}
\caption{The growth history of the \BBS. {\bf a.} The star-formation histories (SFHs) of the three galaxies of the system, inferred from spectrophotometry fitting.
The dotted curve in the z5BBG panel is the SFH derived by \cite{Strait2023ApJ} using a different fitting code.
As their NIRSpec spectra imply, the two closely paired galaxies ELG1+ELG2 are both now undergoing intensive bursts of star formation that started approximately 10 Myr ago, while z5BBG has experienced a short burst in the past and is seeing sudden quenching in the last 10 Myr.
{\bf b.} The stellar mass growth history of the system, reconstructed from the SFHs. The total stellar mass has increased by 0.41 dex in the last 100 Myr, which indicates that $61^{+5}_{-6}\ \%$ of the current stellar mass has formed during the interaction-induced bursts of star formation.
}\label{fig:SFHs}
\end{figure*}

With results of the fits in hand, we applied aperture corrections and lens magnification corrections to obtain the (intrinsic) total physical quantities of the system. We used the latest lens model in this field by G.~Desprez et al.\ (in prep.; summarized in Appendix~\ref{appendix:lens}) to estimate the lens magnification factors ($\mu_{\rm best}$) at the position of each source on the sky.
The aperture correction factor was taken from \cite{Asada2023MNRAS}, where the normalization factor from $0.\!\!^{\prime\prime}3$-diameter aperture to total photometry was estimated based on the curve of growth in NIRCam images.
The best-fit models are presented in Figure \ref{fig:Prospector_SEDs} along with the observational data, and the inferred SFHs normalized with the peak SFRs are shown in Figure \ref{fig:SFHs}. That figure also plots the SFH of the z5BBG galaxy estimated by \cite{Strait2023ApJ} with a different fitting code (\texttt{Bagpipes}; red dotted curve), showing that it agrees well with our {\tt Prospector} result. Table~\ref{tab:Phys_params} lists the inferred physical properties of the three members of the BBS. All three members are low-mass galaxies, ranging in stellar mass from $\sim10^6 M_\odot$ to $\sim10^{7.5} M_\odot$. Although the $z\sim5$ star forming main sequence (SFMS) is not constrained at the extremely low stellar masses of ELG2 and ELG2, these two galaxies are consistent with being above the SFMS when we extrapolate the \citet{Merida2026sfms} SFMS relation to these low masses.  Meanwhile, z5BBG is below the SFMS, as might also be expected from its spectral features.

\section{Results and Discussion}\label{sec:results}

As mentioned  in Sec~\ref{sec:spectral-analysis}, both ELG1 and ELG2 (blue and cyan in Figure \ref{fig1}c)  show extremely strong rest optical emission lines such as \Ha, \Hb, \Hg, and \OIII\ (with H$\alpha$ EW of 3000$\pm$100 \AA\ and 2500$\pm$400 \AA, respectively), with a Balmer jump continuum break at $\lambda_{\rm obs}\sim2.4\ \mu$m.
These spectral features are typical of galaxies experiencing a recently-started burst of star formation \citep{Cameron2024MNRAS,Roberts-Borsani2024ApJ,Mowla2024Natur}.
The strong \Ha\ lines fall in the NIRCam F410M and F444W filters, giving these galaxies their red colors in the color composite images in Figure~\ref{fig1}.
In contrast, as also reported in \cite{Strait2023ApJ},  z5BBG (red spectrum in Figure \ref{fig1}c) only shows a very weak \Ha\ line with an evident Balmer break, indicative of the recent cessation of star formation within this galaxy.
The positive continuum break at $\lambda_{\rm obs}\sim2.4\ \mu$m makes z5BBG appear yellow-white in the color composite images.

\subsection{Stellar mass growth}\label{sec:Discussion_SFHs}

The star formation histories (SFHs) inferred by spectrophotometry fitting in  Sec.~\ref{sec:SEDfitting} reflect the extreme nature of the \BBS.  As Figure \ref{fig:SFHs}a shows, 
ELG1 and ELG2 are both experiencing intense bursts of star formation that started in the last 10 Myr, likely triggered by the interaction event between these two galaxies.  In contrast, z5BBG has experienced a rise in star formation a few tens of Myr earlier but is now undergoing  precipitous quenching. The previous burst of star formation that z5BBG experienced in the recent past started $\lesssim100$ Myr ago, which is a timescale comparable to the crossing time of the ELG1+ELG2 pair from z5BBG, estimated to be $\sim$20 Myr from the velocity difference and projected separation. We thus surmise that the previous burst of star formation in z5BBG is connected to this galaxy's past interaction with either ELG1, ELG2, or both.
Therefore, it appears that all three interacting galaxies experienced fast star formation bursts with timescales of $<100$Myr initiated by galaxy-galaxy interactions. The two galaxies in the close pair, ELG1 and ELG2, are in the rising phase of their bursts, while z5BBG is at the final phase of its burst.
Such interaction-triggered fast starbursts have been implied particularly in low-mass high-$z$ galaxies both by observations \citep{Asada2024MNRAS,Witten2024NatAs} and simulations \citep{Dome2024MNRAS}, and the \BBS\ appears to be a highly magnified poster-child of this process. 

The total stellar mass growth history of the \BBS, reconstructed from the SFHs and shown in Figure \ref{fig:SFHs}b, shows the importance of the interaction-induced starbursts in early hierarchical galaxy assembly.
As we found in Sec.~\ref{sec:SEDfitting}, the lens-corrected stellar masses of ELG1, ELG2, and z5BBG are $\log(M_*/M_\odot)=6.46^{+0.03}_{-0.02}$, $6.03^{+0.08}_{-0.05}$, $7.54^{+0.01}_{-0.01}$, respectively, and thus the total stellar mass of the system is $\log(M_*/M_\odot)=7.57\pm0.02$ at the epoch of  observation.
The total stellar mass 100 Myr earlier, before the galaxies experienced their recent bursts, was $\log(M_*/M_\odot)=7.16\pm0.06$, and thus the total stellar mass has increased by a factor of $\times2.6\pm0.5$ in the last 100 Myr.
This means that at least $61^{+5}_{-6}\ \%$ of the current stellar mass in this system formed during the interaction-triggered starbursts, suggesting the stellar mass growth in low-mass high-z galaxies is considerably accelerated by interactions.
Indeed, the  mass growth enhancement could be even higher given that the burst episodes in ELG1 and ELG2 have not yet ended and are expected to form additional stars in the next few 10 Myrs.%

\subsection{Possible future descendants}\label{sec:Discussion_AbundanceMatching}

\begin{figure}[t]
\centering
    \includegraphics[width=1\linewidth]
{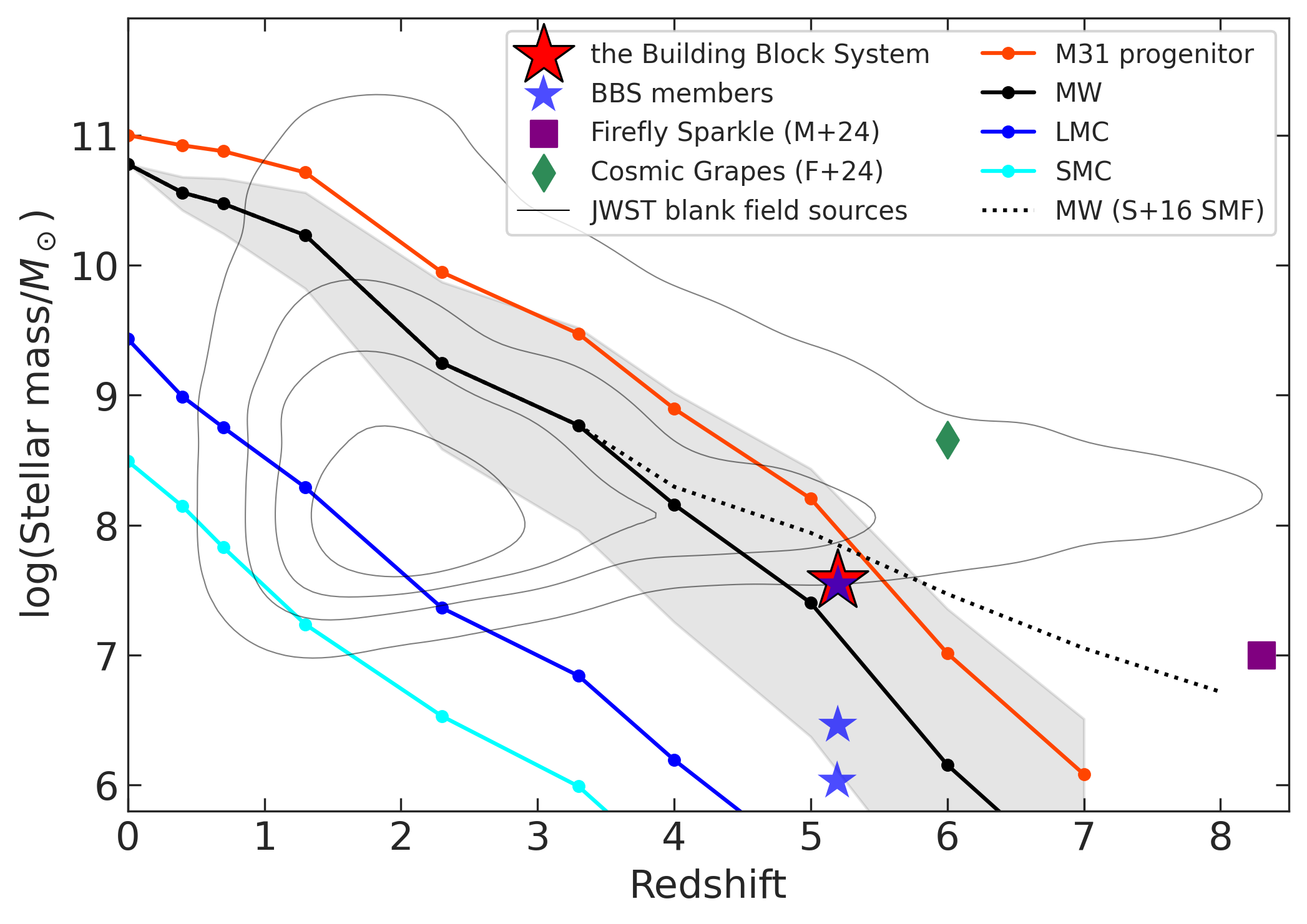}
\caption{The \BBS\ as a MW-like galaxy progenitor at $z=5.2$. Progenitor masses over cosmic time for a given stellar mass at $z=0$ are computed based on the abundance matching technique \cite{Behroozi2013ApJ} with stellar mass functions (SMFs) by \cite{Baldry2012MNRAS,Weaver2023AAS}, shown with colored lines. We computed the progenitor masses of SMC (cyan), LMC (blue), the Milky Way (black), and M31 (red), and the gray shaded area denotes the 16th-to-84th percentile of progenitor masses of the MW. The black dotted line shows MW progenitors assuming a different SMF \cite{Song2016ApJ} at $z>3$. The red filled star presents the current total mass of the \BBS, with the small blue stars marking the three member galaxies. For comparison, we also plot other highly lensed sources with recent JWST NIRSpec observations (purple square, \citealt{Mowla2024Natur}; green diamond, \citealt{Fujimoto2024arxiv}), and the photometric source distribution with high enough S/N to study the internal physics obtained in a JWST blank field survey (gray contours). 
}\label{fig:Progenitor}
\end{figure}

The components of the \BBS, observed in the process of interaction and (likely) merging $\sim$1~Gyr after the Big Bang, are low-mass objects. But what may they evolve into in the future?  To answer this question, we need to consider the expected $z\sim 5$ progenitor masses of different types of present-day galaxies, and to do so we used a semi-analytical approach similar to that of  \citet{Mowla2024Natur} and \citet{Tan2024arXiv}.  Specifically, we  used the abundance matching code from \cite{Behroozi2013ApJ} to compute the redshift evolution of the co-moving number density of a galaxy population from $z=0$ to a given redshift $z=z_1$. 
\cite{Behroozi2013ApJ} matched galaxies and dark matter halos in a simulation using their cumulative number densities, and tracked the halo growth history taking into account the effects of mergers and scatters of mass-accretion histories.
The \citet{Behroozi2013ApJ} code allows us to  compute the cumulative number density of a galaxy population at $z=z_1$ if we provide the initial cumulative number density of the population at $z=z_0$.

We used galaxy stellar mass functions from the literature \citep{Baldry2012MNRAS,Weaver2023AAS} to convert the cumulative galaxy number density into the galaxy stellar mass at each redshift.  We then computed the progenitor masses of several well-known present-day galaxies that span wide range of stellar masses in order to examine how massive the galaxies we observed at $z=5.2$ may become at $z=0$.
Figure \ref{fig:Progenitor} shows the resulting progenitor stellar masses of M31 (red), Milky Way (black), LMC (blue), and SMC (cyan), with the 16th- to 84-th percentile of progenitor stellar masses of MW-mass galaxies  shown with a gray band. The observed stellar mass of the \BBS\ is consistent with being the progenitor mass of present-day $M_\star\sim 10^{11}M_\odot$ massive galaxies such as the Milky Way (MW) or M31.  

The progenitor mass estimation inherently depends on the stellar mass function assumption. To examine the effect of this assumption, we also track the MW-like galaxy progenitors using a different SMF at $z>4$ by \cite{Song2016ApJ}, and show the result with the black dotted line in Figure \ref{fig:Progenitor}.
This different SMF assumption affects our analysis only within the 1$\sigma$ uncertainty up to $z\sim6$, and the effect on our conclusions is therefore negligible. 

On the basis of the abundance-matching analysis presented above, we conclude that the \BBS\ is consistent with being the progenitor of a present-day massive galaxy. Whether BBS will evolve into such a massive system, or whether it represents the main, most massive branch of the merger tree that leads to such a massive galaxy, is of course unknown for this specific object. However, the analysis presented here suggests that such fate is a possible outcome for \BBS\ in its future. 

Figure \ref{fig:Progenitor} also highlights the power of the gravitational lensing effect in studying progenitors of MW-like galaxies at $z>5$.
Typical deep blank-field surveys with NIRCam would provide photometric samples down to $\sim10^{7.5}\ M_\odot$, and most of the MW-like galaxy progenitors at $z>5$ are below this selection limit.
The gray contour in Figure \ref{fig:Progenitor} shows the photometric source distribution in the $M_\star$-$z$ diagram, selected as S/N$>10$ from the CANUCS NIRCam Flanking Field observations (effective detection limits are $\sim30.2$ mag at 3-sigma in F444W; \citetalias{Sarrouh2026}), and the median of the MW galaxy progenitors (black curve) falls below the contour at $z\sim4.8$.
This stellar mass limit gets even higher when the internal structure is to be studied; e.g., \citet{Tan2024arXiv} selected their sample requiring S/N$>30$ to explore the spatially resolved properties of MW-like galaxy progenitors up to $z=5$ from NIRCam observations.  Gravitational lensing is thus crucially needed to study the detailed and internal physics behind  early galaxy evolution, as it means we can obtain the needed higher-S/N data at higher spatial resolution.

\subsection{In-situ growth vs.\ hierarchical assembly}

The \BBS\ stands in contrast to recent reports of \emph{in-situ} star cluster formation in high-$z$ lensed galaxies \citep[e.g.,][]{Adamo2024Natur,Fujimoto2024arxiv,Mowla2024Natur,Nakane2025arXiv}.
These recent studies
suggest that clumpy star formation, perhaps caused by gas disk instabilities, can be a major evolutionary process at high-$z$. 
This is in contrast to the \BBS, which is undergoing hierarchical assembly boosted by interaction-triggered star-forming bursts within the merging components. A recent study by \citet{Omori2026arXiv} also shows that nearly half of $z_{\rm spec}\sim5$ galaxies have closely-interacting companions, and their stellar mass growth seems to be higher than pure merging of previously existing stellar masses of their components. Therefore, the classical hierarchical growth picture boosted by interaction-triggered star-formation seems to play a key role in early galaxy assembly.

Given limited number statistics, we cannot conclude which mechanism (\textit{in-situ} clump formation or burst-enhanced hierarchical assembly) dominates. It is even possible that both mechanisms play a role in the early history of present-day massive galaxies, potentially even operating sequentially within the same galaxy.
Nevertheless, the \BBS\ illustrates that hierarchical merging not only remains a viable major process of the early assembly of MW-like galaxies, and that such assembly can be dramatically accelerated, by bursts of star formation, compared to the classical hierarchical paradigm.

\section{Conclusion}\label{sec:concl}

In this paper we analyzed the spectro-photometric properties of the \BBS, a trio of low-mass ($\log (M_\star/M_\odot) \sim 10^{6.03}, 10^{6.46}$, and $10^{7.54}$) $z_{\rm spec}=5.20$ galaxies that are located in close proximity to each other both on the sky and in velocity. Unlike only with close \emph{projected} pairs, the close \emph{physical} proximity of the \BBS's components virtually guarantees that they are interacting, and likely to merge in near future.
The strong gravitational magnification of this system makes it an excellent example of low-mass interacting systems that are likely to evolve into a MW-like mass galaxy by $z=0$.

Our analysis of this system suggests that interactions between low-mass high-$z$ galaxies can produce short bursts of star formation with durations of $\sim$10-100~Myr, and that these star-forming bursts produce significant additional mass over and above the mass of the pre-burst stellar populations.  In the \BBS, the observed bursts boost the mass growth 2.6 $\pm$0.5 times that expected in straightforward ``dry" merging of the existing stellar masses.
This result signifies the importance of galaxy-galaxy interactions in early galaxy assembly, in line with the classical hierarchical paradigm but in contrast to recent high-$z$ lensed galaxy observations with JWST.
Analysis of larger samples are needed to determine the relative contribution of these two mechanisms --- in-situ star-cluster formation vs.\ merging ---  to the early assembly of massive galaxy progenitors. Meanwhile detailed, spatially-resolved  spectroscopic studies of key individual systems, such as those enabled by NIRSpec/IFU spectroscopy, will shed light on the physics at play.

\begin{acknowledgments}
This work is based on observations made with the NASA/ESA/CSA JWST. The data were obtained from the Mikulski Archive for Space Telescopes at the Space Telescope Science Institute, which is operated by the Association of Universities for Research in Astronomy, Inc., under NASA contract NAS5-03127 for JWST.
This research was enabled by Canadian Space Agency grants 18JWST-GTO1 and 24JWGO3A10 and Natural Sciences and Engineering Research Council (NSERC) of Canada grants RGPIN-2020-06023,  RGPAS-2020-00065, and RGPIN-2026-08047 to MS.
YA was supported by JSPS KAKENHI Grant Number 23H00131, and acknowledges ongoing support from the Dunlap Institute for Astronomy \& Astrophysics. The Dunlap Institute is funded through an endowment established by the David Dunlap family and the University of Toronto.
MB, JJ, GR, NM, AH, GF, and VM acknowledge support from the ERC Grant FIRSTLIGHT, Slovenian national research agency ARIS through grants N1-0238 and P1-0188, and ESA PRODEX grant. This research used the Canadian Advanced Network For Astronomy Research (CANFAR) operated in partnership with the Canadian Astronomy Data Centre and The Digital Research Alliance of Canada, with support from the National Research Council of Canada, the Canadian Space Agency, CANARIE, and the Canada Foundation for Innovation. 
\end{acknowledgments}




%
\facilities{HST (WFC3), JWST (NIRCam and NIRSpec)}

\software{astropy \citep{2013A&A...558A..33A,2018AJ....156..123A,2022ApJ...935..167A}, photutils \citep{photutils1.12.0}, msaexp \citep{Brammer2022zndo}
          }


\appendix

\section{Custom NIRSpec Background Subtraction}\label{appendix:spec}

\begin{figure*}[t]
\centering
\includegraphics[width=0.95\textwidth]{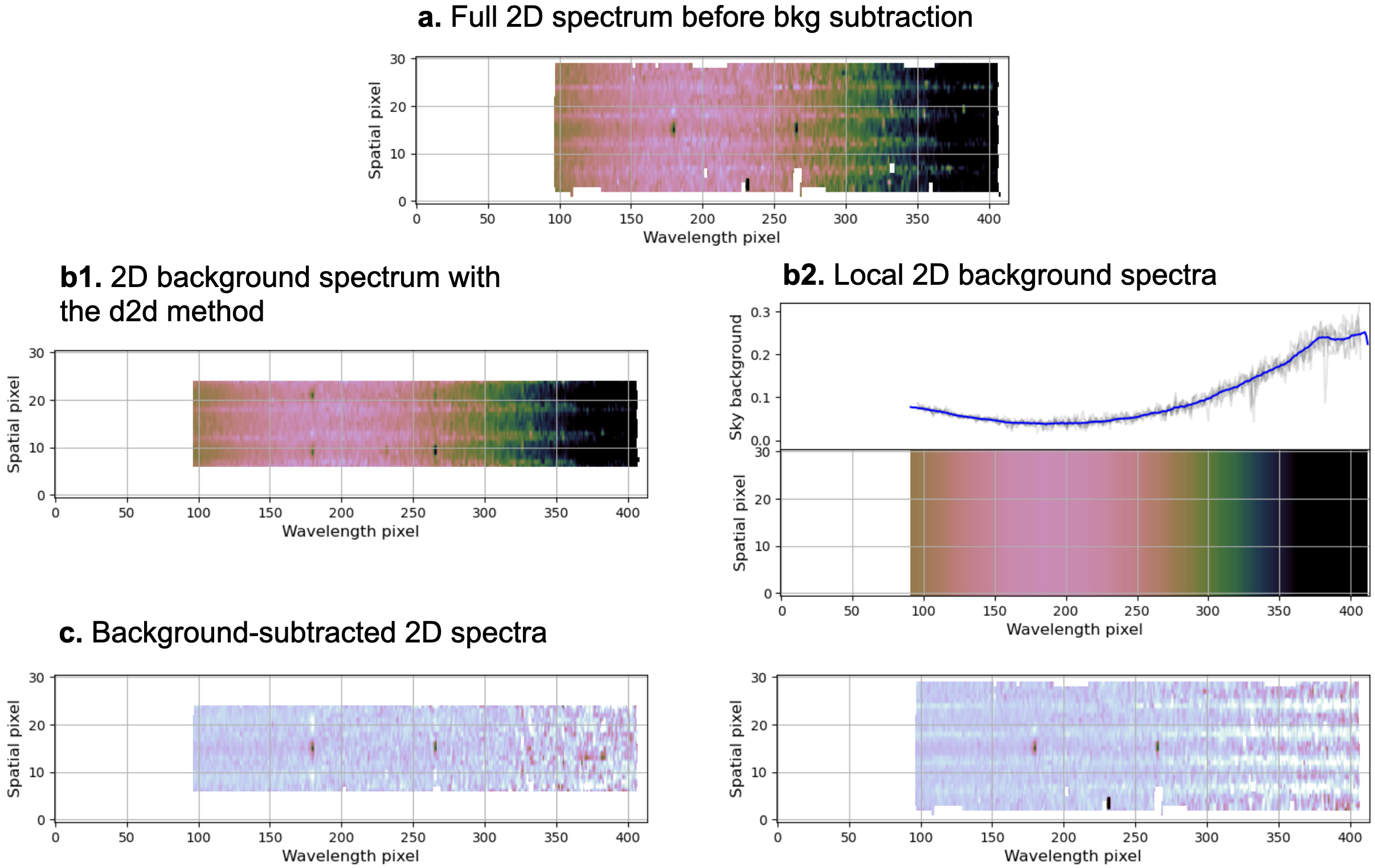}
\caption{The background subtraction process with the standard d2d subtraction (left column) and the local 2D background subtraction (right column).  \textbf{a}, The full 2D spectrum before background subtraction. Both methods start from this 2D spectrum. \textbf{b1}. The d2d background spectrum. The central shutter of this spectrum is used as the background for the source located in the central shutter. \textbf{b2}. The local 2D background spectrum. The upper sub-panel shows the 1D spectrum of the sky background. Gray lines show the individual 1D spectra in the empty pixels, and the blue curve presents the stacked and smoothed 1D sky spectrum. We constructed the 2D background spectrum using this stacked/smoothed 1D spectrum, which is shown in the lower sub-panel. \textbf{c}. Background-subtracted 2D spectra. Left and right panels show the results with the standard d2d and the local 2D background method, respectively.}\label{fig:Bkgsub}
\end{figure*}

One way to perform the background subtraction in an NIRSpec MSA spectrum is by drizzling the 2D spectrum by 6 pixels in the cross-dispersion direction (which corresponds to the length of a shutter), combining the drizzled 2D spectra to obtain the background spectrum in the central shutter, and subtracting it from the original 2D spectrum (e.g., \cite{Strait2023ApJ,DEugenio2024AAP}). This is the approach we followed for the spectrum of z5BBG, as described in \cite{Strait2023ApJ}. This subtraction strategy assumes the adjacent shutters do not contain any sources and can therefore be used as background shutters. However, in the case of the ELG1+ELG2 galaxy pair, ELG1 was located in the central shutter and ELG2 was in the adjacent top shutter (Figure \ref{fig:ELG12_2Dspec}), and consequently we cannot use the standard drizzle 2D background subtraction (which we refer to as d2d subtraction hereafter). To deal with this problem, we developed a custom background subtraction method which creates a local 2D background spectrum by stacking and smoothing the 1D spectra taken from empty regions along the slit direction. Figure \ref{fig:Bkgsub} illustrates the two background subtraction methods: the standard d2d method on the left, and our custom method on the right. In the standard d2d subtraction method, the background spectrum of the central shutter was obtained by drizzling the full 2D spectrum by 6 pixels (panel b1 in the figure). By subtracting this d2d spectrum from the full 2D spectrum (panel a), one derives the background-subtracted 2D spectrum (panel c left). On the other hand, in our local 2D background subtraction method, we first obtain the 1D sky spectrum from the full 2D spectrum before the subtraction. For this, we use the 1D spectra from empty pixels in the adjacent shutters (grey curves in panel b2 top), and then median-stack and smooth these 1D spectra to obtain the 1D sky spectrum (blue curve in b2 top). Assuming the background spectrum to be spatially uniform, we thereby obtain the local 2D background spectrum (panel b2 bottom). By subtracting this 2D spectrum from the full 2D spectrum, we derive the background-subtracted 2D spectrum with this method (panel c right).

\begin{figure}[t]
\centering
    \includegraphics[width=0.75\linewidth]
{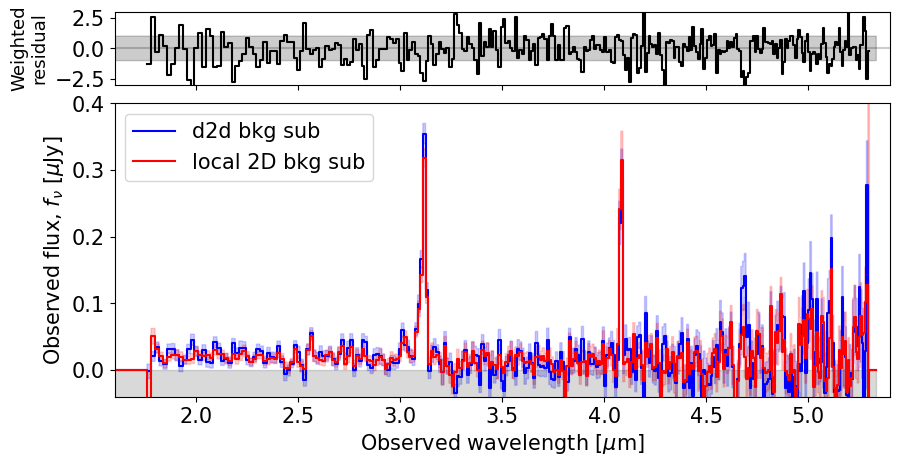}
\caption{Background-subtracted 1D spectra of CANUCS-1111496. Red and blue show the resulting 1D spectra with the standard d2d background subtraction and the local 2D background subtraction method used in this work, respectively. The two methods give the consistent 1D spectra. The top sub-panel shows the residual of the two spectra, weighted by the uncertainty. The gray shaded area in the panel marks $\pm1\sigma$.}\label{fig:Spec_comp_d2d}
\end{figure}

Before using our custom method to analyze the spectrum of ELG1+ELG2, we first demonstrate and test this background subtraction method on another galaxy, CANUCS-1111496. This galaxy is at a similar redshift to the ELG1+ELG2 galaxy pair but has no sources contaminating its upper and lower shutters, which makes it possible to apply both background subtraction methods to its spectrum and compare the results. Applying both the d2d and our local 2D subtraction methods, we generated two sets of background-subtracted spectra of CANUCS-1111496. We then extracted individual 1D spectra by collapsing the 2D spectra with an inverse-variance weighted kernel \citep{Horne1986PASP}. Figure \ref{fig:Spec_comp_d2d} compares the resulting 1D spectra from the standard d2d background subtraction method (blue) and the local 2D background subtraction method (red). The figure demonstrates that the local 2D background subtraction method gives a result consistent with that of the d2d subtraction. Moreover, in cases where the adjacent shutters contain other sources (as is the case for ELG1+ELG2 in this work), the standard d2d subtraction strategy cannot work properly, while our local 2D subtraction avoids the self-subtraction issue and enables us to derive the spectrum not only from the source in the central shutter but also from the source in the adjacent shutter. Indeed, Figure \ref{fig:ELG12_2Dspec} shows the 2D spectrum of the ELG1+ELG2 system, where we were able to derive both ELG1 and ELG2 spectra separately while avoiding self-subtraction.  

As an additional benefit, the local 2D background subtraction strategy can reduce the noise level in the resulting 1D spectrum, as can be seen in Figure \ref{fig:Spec_comp_d2d}, where the continuum spectrum is smoother in the result (red) when compared with the standard d2d approach (blue). This noise reduction arises because of the stacking and smoothing that suppresses the noise level in the background spectrum as compared to that in the standard d2d approach, which does not benefit from this stacking of the background signal. 

The 2D spectrum from the local 2D background subtraction can be affected by bar shadows, particularly in the red end of the wavelength range (Figure \ref{fig:Bkgsub}c right), as compared to the d2d subtraction (panel c left). Although the effect of bar shadows is corrected in the path-loss correction step using the standard STScI pipeline, some bar shadow residuals remain after level 1 processing (Figure \ref{fig:Bkgsub}a). These residuals are included as the background in the standard d2d method (panel b1), where they are canceled out in the background-subtracted 2D spectrum (panel c left). In contrast, the bar shadow residuals remain in the local 2D subtraction spectrum (panel c right).  However, when the source of interest is well centered in a shutter, the effect of bar shadows on the resulting 1D spectrum is negligible: we cannot see the systematic effect due to the bar shadows in the spectrum of CANUCS-1111496 (Figure \ref{fig:Spec_comp_d2d}). But, when the source is particularly close to the edge of the shutter, the local 2D background subtraction method can over-subtract the sky background at long wavelengths due to the bar shadows at $\lambda_{\rm obs}\gtrsim4.3\ \mu$m, an effect that can be seen in the spectrum of ELG2 (Figure \ref{fig:ELG12_1Dspec}b). When this happens, the continuum shape at long wavelengths should be treated with caution when using the local 2D background subtraction method. 

\section{Gravitational lens modeling}\label{appendix:lens}

The strong lensing model used in this work is similar to the one used in \cite{Estrada-carpenter2024} and is fully described in Desprez et al.\ (in prep.). It is a model built using the code {\tt Lenstool} \citep{Kneib1993,Jullo2007}, which describes the cluster mass distribution as a collection of halos parametrized by double Pseudo-Isothermal Elliptical (dPIE \cite{Eliasdottir2007arXiv}) profiles. Different sizes of dPIE halos are considered to model the overall cluster contribution to the mass model as well as the individual cluster-member ones. These halo properties are constrained leveraging the observed position and redshift of multiple-image systems identified in the fields from \cite{Mahler2019,Jauzac2019}, but also from the CANUCS images and spectra \citetalias{Sarrouh2026}. Only the gold standard images, as prescribed by \cite{Gledhill2024,Rihtarsic2025}, are used in the model  optimization, amounting for a total of 82 images or clumps from 19 multiple systems. The double images of ELG1, ELG2, and z5BBG are included as constraints in the model, and the best fit model manages to predict the positions of these systems with an average distance from their observed positions of $0.\!\!^{\prime\prime}04$, $0.\!\!^{\prime\prime}05$, and $0.\!\!^{\prime\prime}2$ respectively. The average absolute offset is $0.\!\!^{\prime\prime}44$ when considering all the multiple images used to constrain the model, indicating that ELG1, ELG2, and z5BBG are among the best fit sources. 




\bibliography{bibliography}{}

@software{photutils1.12.0,
       author = {{Bradley}, Larry and {Sip{\H{o}}cz}, Brigitta and {Robitaille}, Thomas and {Tollerud}, Erik and {Vin{\'\i}cius}, Z{\'e} and {Deil}, Christoph and {Barbary}, Kyle and {Wilson}, Tom J and {Busko}, Ivo and {Donath}, Axel and {G{\"u}nther}, Hans Moritz and {Cara}, Mihai and {Lim}, P.~L. and {Me{\ss}linger}, Sebastian and {Burnett}, Zach and {Conseil}, Simon and {Droettboom}, Michael and {Bostroem}, Azalee and {Bray}, E.~M. and {Andersen Bratholm}, Lars and {Jamieson}, William and {Ginsburg}, Adam and {Barentsen}, Geert and {Craig}, Matt and {Pascual}, Sergio and {Rathi}, Shivangee and {Perrin}, Marshall and {Morris}, Brett M. and {Perren}, Gabriel},
        title = "{astropy/photutils: 1.12.0}",
         year = 2024,
        month = apr,
          eid = {10.5281/zenodo.10967176},
          doi = {10.5281/zenodo.10967176},
      version = {1.12.0},
    publisher = {Zenodo},
       adsurl = {https://ui.adsabs.harvard.edu/abs/2024zndo..10967176B}
}

@ARTICLE{2022ApJ...935..167A,
       author = {{Astropy Collaboration} and {Price-Whelan}, Adrian M. and {Lim}, Pey Lian and {Earl}, Nicholas and {Starkman}, Nathaniel and {Bradley}, Larry and {Shupe}, David L. and {Patil}, Aarya A. and {Corrales}, Lia and {Brasseur}, C.~E. and {N{\"o}the}, Maximilian and {Donath}, Axel and {Tollerud}, Erik and {Morris}, Brett M. and {Ginsburg}, Adam and {Vaher}, Eero and {Weaver}, Benjamin A. and {Tocknell}, James and {Jamieson}, William and {van Kerkwijk}, Marten H. and {Robitaille}, Thomas P. and {Merry}, Bruce and {Bachetti}, Matteo and {G{\"u}nther}, H. Moritz and {Aldcroft}, Thomas L. and {Alvarado-Montes}, Jaime A. and {Archibald}, Anne M. and {B{\'o}di}, Attila and {Bapat}, Shreyas and {Barentsen}, Geert and {Baz{\'a}n}, Juanjo and {Biswas}, Manish and {Boquien}, M{\'e}d{\'e}ric and {Burke}, D.~J. and {Cara}, Daria and {Cara}, Mihai and {Conroy}, Kyle E. and {Conseil}, Simon and {Craig}, Matthew W. and {Cross}, Robert M. and {Cruz}, Kelle L. and {D'Eugenio}, Francesco and {Dencheva}, Nadia and {Devillepoix}, Hadrien A.~R. and {Dietrich}, J{\"o}rg P. and {Eigenbrot}, Arthur Davis and {Erben}, Thomas and {Ferreira}, Leonardo and {Foreman-Mackey}, Daniel and {Fox}, Ryan and {Freij}, Nabil and {Garg}, Suyog and {Geda}, Robel and {Glattly}, Lauren and {Gondhalekar}, Yash and {Gordon}, Karl D. and {Grant}, David and {Greenfield}, Perry and {Groener}, Austen M. and {Guest}, Steve and {Gurovich}, Sebastian and {Handberg}, Rasmus and {Hart}, Akeem and {Hatfield-Dodds}, Zac and {Homeier}, Derek and {Hosseinzadeh}, Griffin and {Jenness}, Tim and {Jones}, Craig K. and {Joseph}, Prajwel and {Kalmbach}, J. Bryce and {Karamehmetoglu}, Emir and {Ka{\l}uszy{\'n}ski}, Miko{\l}aj and {Kelley}, Michael S.~P. and {Kern}, Nicholas and {Kerzendorf}, Wolfgang E. and {Koch}, Eric W. and {Kulumani}, Shankar and {Lee}, Antony and {Ly}, Chun and {Ma}, Zhiyuan and {MacBride}, Conor and {Maljaars}, Jakob M. and {Muna}, Demitri and {Murphy}, N.~A. and {Norman}, Henrik and {O'Steen}, Richard and {Oman}, Kyle A. and {Pacifici}, Camilla and {Pascual}, Sergio and {Pascual-Granado}, J. and {Patil}, Rohit R. and {Perren}, Gabriel I. and {Pickering}, Timothy E. and {Rastogi}, Tanuj and {Roulston}, Benjamin R. and {Ryan}, Daniel F. and {Rykoff}, Eli S. and {Sabater}, Jose and {Sakurikar}, Parikshit and {Salgado}, Jes{\'u}s and {Sanghi}, Aniket and {Saunders}, Nicholas and {Savchenko}, Volodymyr and {Schwardt}, Ludwig and {Seifert-Eckert}, Michael and {Shih}, Albert Y. and {Jain}, Anany Shrey and {Shukla}, Gyanendra and {Sick}, Jonathan and {Simpson}, Chris and {Singanamalla}, Sudheesh and {Singer}, Leo P. and {Singhal}, Jaladh and {Sinha}, Manodeep and {Sip{\H{o}}cz}, Brigitta M. and {Spitler}, Lee R. and {Stansby}, David and {Streicher}, Ole and {{\v{S}}umak}, Jani and {Swinbank}, John D. and {Taranu}, Dan S. and {Tewary}, Nikita and {Tremblay}, Grant R. and {de Val-Borro}, Miguel and {Van Kooten}, Samuel J. and {Vasovi{\'c}}, Zlatan and {Verma}, Shresth and {de Miranda Cardoso}, Jos{\'e} Vin{\'\i}cius and {Williams}, Peter K.~G. and {Wilson}, Tom J. and {Winkel}, Benjamin and {Wood-Vasey}, W.~M. and {Xue}, Rui and {Yoachim}, Peter and {Zhang}, Chen and {Zonca}, Andrea and {Astropy Project Contributors}},
        title = "{The Astropy Project: Sustaining and Growing a Community-oriented Open-source Project and the Latest Major Release (v5.0) of the Core Package}",
      journal = {\apj},
         year = 2022,
        month = aug,
       volume = {935},
       number = {2},
          eid = {167},
        pages = {167},
          doi = {10.3847/1538-4357/ac7c74},
archivePrefix = {arXiv},
       eprint = {2206.14220},
 primaryClass = {astro-ph.IM},
       adsurl = {https://ui.adsabs.harvard.edu/abs/2022ApJ...935..167A}
}

@ARTICLE{2018AJ....156..123A,
       author = {{Astropy Collaboration} and {Price-Whelan}, A.~M. and {Sip{\H{o}}cz}, B.~M. and {G{\"u}nther}, H.~M. and {Lim}, P.~L. and {Crawford}, S.~M. and {Conseil}, S. and {Shupe}, D.~L. and {Craig}, M.~W. and {Dencheva}, N. and {Ginsburg}, A. and {VanderPlas}, J.~T. and {Bradley}, L.~D. and {P{\'e}rez-Su{\'a}rez}, D. and {de Val-Borro}, M. and {Aldcroft}, T.~L. and {Cruz}, K.~L. and {Robitaille}, T.~P. and {Tollerud}, E.~J. and {Ardelean}, C. and {Babej}, T. and {Bach}, Y.~P. and {Bachetti}, M. and {Bakanov}, A.~V. and {Bamford}, S.~P. and {Barentsen}, G. and {Barmby}, P. and {Baumbach}, A. and {Berry}, K.~L. and {Biscani}, F. and {Boquien}, M. and {Bostroem}, K.~A. and {Bouma}, L.~G. and {Brammer}, G.~B. and {Bray}, E.~M. and {Breytenbach}, H. and {Buddelmeijer}, H. and {Burke}, D.~J. and {Calderone}, G. and {Cano Rodr{\'\i}guez}, J.~L. and {Cara}, M. and {Cardoso}, J.~V.~M. and {Cheedella}, S. and {Copin}, Y. and {Corrales}, L. and {Crichton}, D. and {D'Avella}, D. and {Deil}, C. and {Depagne}, {\'E}. and {Dietrich}, J.~P. and {Donath}, A. and {Droettboom}, M. and {Earl}, N. and {Erben}, T. and {Fabbro}, S. and {Ferreira}, L.~A. and {Finethy}, T. and {Fox}, R.~T. and {Garrison}, L.~H. and {Gibbons}, S.~L.~J. and {Goldstein}, D.~A. and {Gommers}, R. and {Greco}, J.~P. and {Greenfield}, P. and {Groener}, A.~M. and {Grollier}, F. and {Hagen}, A. and {Hirst}, P. and {Homeier}, D. and {Horton}, A.~J. and {Hosseinzadeh}, G. and {Hu}, L. and {Hunkeler}, J.~S. and {Ivezi{\'c}}, {\v{Z}}. and {Jain}, A. and {Jenness}, T. and {Kanarek}, G. and {Kendrew}, S. and {Kern}, N.~S. and {Kerzendorf}, W.~E. and {Khvalko}, A. and {King}, J. and {Kirkby}, D. and {Kulkarni}, A.~M. and {Kumar}, A. and {Lee}, A. and {Lenz}, D. and {Littlefair}, S.~P. and {Ma}, Z. and {Macleod}, D.~M. and {Mastropietro}, M. and {McCully}, C. and {Montagnac}, S. and {Morris}, B.~M. and {Mueller}, M. and {Mumford}, S.~J. and {Muna}, D. and {Murphy}, N.~A. and {Nelson}, S. and {Nguyen}, G.~H. and {Ninan}, J.~P. and {N{\"o}the}, M. and {Ogaz}, S. and {Oh}, S. and {Parejko}, J.~K. and {Parley}, N. and {Pascual}, S. and {Patil}, R. and {Patil}, A.~A. and {Plunkett}, A.~L. and {Prochaska}, J.~X. and {Rastogi}, T. and {Reddy Janga}, V. and {Sabater}, J. and {Sakurikar}, P. and {Seifert}, M. and {Sherbert}, L.~E. and {Sherwood-Taylor}, H. and {Shih}, A.~Y. and {Sick}, J. and {Silbiger}, M.~T. and {Singanamalla}, S. and {Singer}, L.~P. and {Sladen}, P.~H. and {Sooley}, K.~A. and {Sornarajah}, S. and {Streicher}, O. and {Teuben}, P. and {Thomas}, S.~W. and {Tremblay}, G.~R. and {Turner}, J.~E.~H. and {Terr{\'o}n}, V. and {van Kerkwijk}, M.~H. and {de la Vega}, A. and {Watkins}, L.~L. and {Weaver}, B.~A. and {Whitmore}, J.~B. and {Woillez}, J. and {Zabalza}, V. and {Astropy Contributors}},
        title = "{The Astropy Project: Building an Open-science Project and Status of the v2.0 Core Package}",
      journal = {\aj},
         year = 2018,
        month = sep,
       volume = {156},
       number = {3},
          eid = {123},
        pages = {123},
          doi = {10.3847/1538-3881/aabc4f},
archivePrefix = {arXiv},
       eprint = {1801.02634},
 primaryClass = {astro-ph.IM},
       adsurl = {https://ui.adsabs.harvard.edu/abs/2018AJ....156..123A}
}

@ARTICLE{2013A&A...558A..33A,
       author = {{Astropy Collaboration} and {Robitaille}, Thomas P. and
         {Tollerud}, Erik J. and {Greenfield}, Perry and {Droettboom}, Michael and
         {Bray}, Erik and {Aldcroft}, Tom and {Davis}, Matt and
         {Ginsburg}, Adam and {Price-Whelan}, Adrian M. and
         {Kerzendorf}, Wolfgang E. and {Conley}, Alexander and {Crighton}, Neil and
         {Barbary}, Kyle and {Muna}, Demitri and {Ferguson}, Henry and
         {Grollier}, Fr{\'e}d{\'e}ric and {Parikh}, Madhura M. and
         {Nair}, Prasanth H. and {Unther}, Hans M. and {Deil}, Christoph and
         {Woillez}, Julien and {Conseil}, Simon and {Kramer}, Roban and
         {Turner}, James E.~H. and {Singer}, Leo and {Fox}, Ryan and
         {Weaver}, Benjamin A. and {Zabalza}, Victor and {Edwards}, Zachary I. and
         {Azalee Bostroem}, K. and {Burke}, D.~J. and {Casey}, Andrew R. and
         {Crawford}, Steven M. and {Dencheva}, Nadia and {Ely}, Justin and
         {Jenness}, Tim and {Labrie}, Kathleen and {Lim}, Pey Lian and
         {Pierfederici}, Francesco and {Pontzen}, Andrew and {Ptak}, Andy and
         {Refsdal}, Brian and {Servillat}, Mathieu and {Streicher}, Ole},
        title = "{Astropy: A community Python package for astronomy}",
      journal = {\aap},
         year = "2013",
        month = "Oct",
       volume = {558},
          eid = {A33},
        pages = {A33},
          doi = {10.1051/0004-6361/201322068},
archivePrefix = {arXiv},
       eprint = {1307.6212},
 primaryClass = {astro-ph.IM},
       adsurl = {https://ui.adsabs.harvard.edu/abs/2013A&A...558A..33A}
}

@ARTICLE{Omori2026arXiv,
       author = {{Omori}, Kiyoaki C. and {Sawicki}, Marcin and {M{\'e}rida}, Rosa M. and {Desprez}, Guillaume and {Abraham}, Roberto and {Brada{\v{c}}}, Maru{\v{s}}a and {Martis}, Nicholas S. and {Muzzin}, Adam and {Noirot}, Ga{\"e}l and {Sarrouh}, Ghassan T. and {Willott}, Christopher J. and {Marchesini}, Danilo and {Myers}, Katherine},
        title = "{Interaction-induced star formation boosts stellar mass assembly in $z\sim5$ galaxies}",
      journal = {arXiv e-prints},
         year = 2026,
        month = jun,
          eid = {arXiv:2606.28590},
        pages = {arXiv:2606.28590},
          doi = {10.48550/arXiv.2606.28590},
archivePrefix = {arXiv},
       eprint = {2606.28590},
 primaryClass = {astro-ph.GA},
       adsurl = {https://ui.adsabs.harvard.edu/abs/2026arXiv260628590O}
}

@ARTICLE{Chemerynska2024ApJ,
       author = {{Chemerynska}, Iryna and {Atek}, Hakim and {Dayal}, Pratika and {Furtak}, Lukas J. and {Feldmann}, Robert and {Greene}, Jenny E. and {Maseda}, Michael V. and {Nanayakkara}, Themiya and {Oesch}, Pascal A. and {Fujimoto}, Seiji and {Labb{\'e}}, Ivo and {Bezanson}, Rachel and {Brammer}, Gabriel and {Cutler}, Sam E. and {Leja}, Joel and {Pan}, Richard and {Price}, Sedona H. and {Wang}, Bingjie and {Weaver}, John R. and {Whitaker}, Katherine E.},
        title = "{The Extreme Low-mass End of the Mass─Metallicity Relation at z {\ensuremath{\sim}} 7}",
      journal = {\apjl},
         year = 2024,
        month = nov,
       volume = {976},
       number = {1},
          eid = {L15},
        pages = {L15},
          doi = {10.3847/2041-8213/ad8dc9},
archivePrefix = {arXiv},
       eprint = {2407.17110},
 primaryClass = {astro-ph.GA},
       adsurl = {https://ui.adsabs.harvard.edu/abs/2024ApJ...976L..15C}
}

@ARTICLE{Atek2024Natur,
       author = {{Atek}, Hakim and {Labb{\'e}}, Ivo and {Furtak}, Lukas J. and {Chemerynska}, Iryna and {Fujimoto}, Seiji and {Setton}, David J. and {Miller}, Tim B. and {Oesch}, Pascal and {Bezanson}, Rachel and {Price}, Sedona H. and {Dayal}, Pratika and {Zitrin}, Adi and {Kokorev}, Vasily and {Weaver}, John R. and {Brammer}, Gabriel and {Dokkum}, Pieter van and {Williams}, Christina C. and {Cutler}, Sam E. and {Feldmann}, Robert and {Fudamoto}, Yoshinobu and {Greene}, Jenny E. and {Leja}, Joel and {Maseda}, Michael V. and {Muzzin}, Adam and {Pan}, Richard and {Papovich}, Casey and {Nelson}, Erica J. and {Nanayakkara}, Themiya and {Stark}, Daniel P. and {Stefanon}, Mauro and {Suess}, Katherine A. and {Wang}, Bingjie and {Whitaker}, Katherine E.},
        title = "{Most of the photons that reionized the Universe came from dwarf galaxies}",
      journal = {\nat},
         year = 2024,
        month = feb,
       volume = {626},
       number = {8001},
        pages = {975-978},
          doi = {10.1038/s41586-024-07043-6},
archivePrefix = {arXiv},
       eprint = {2308.08540},
 primaryClass = {astro-ph.GA},
       adsurl = {https://ui.adsabs.harvard.edu/abs/2024Natur.626..975A}
}

@ARTICLE{Asada2026arXiv,
       author = {{Asada}, Yoshihisa and {Fujimoto}, Seiji and {Chisholm}, John and {Naidu}, Rohan P. and {Atek}, Hakim and {Brammer}, Gabriel and {Furtak}, Lukas J. and {Kokorev}, Vasily and {Pan}, Richard and {Basu}, Arghyadeep and {Bromm}, Volker and {Dessauges-Zavadsky}, Miroslava and {Hsiao}, Tiger Yu-Yang and {Jecmen}, Michelle and {Korber}, Damien and {Liu}, Boyuan and {McKinney}, Jed and {McQuinn}, Kristen B.~W. and {Schaerer}, Daniel},
        title = "{GLIMPSE-DDT spectroscopic properties of faint-end galaxies at $z\sim6$: Towards first metal enrichment, dust production, and ionizing photon production}",
      journal = {arXiv e-prints},
         year = 2026,
        month = jan,
          eid = {arXiv:2601.20045},
        pages = {arXiv:2601.20045},
          doi = {10.48550/arXiv.2601.20045},
archivePrefix = {arXiv},
       eprint = {2601.20045},
 primaryClass = {astro-ph.GA},
       adsurl = {https://ui.adsabs.harvard.edu/abs/2026arXiv260120045A}
}

@ARTICLE{Nakane2025arXiv,
       author = {{Nakane}, Minami and {Kokorev}, Vasily and {Fujimoto}, Seiji and {Ouchi}, Masami and {McLeod}, Derek J. and {Golubchik}, Miriam and {Oguri}, Masamune and {Zitrin}, Adi and {Bondestam}, Cecilia and {Donnan}, Callum T. and {Brammer}, Gabriel and {Finkelstein}, Steven L. and {Willott}, Chris and {Adamo}, Angela and {Vanzella}, Eros and {Brada{\v{c}}}, Marusa and {Messa}, Matteo and {Yanagisawa}, Hiroto and {Sun}, Fengwu and {Ferguson}, Henry C. and {Lucas}, Ray A. and {Coe}, Dan and {Richard}, Johan and {Abdurro'uf} and {Akins}, Hollis B. and {Allingham}, Joseph F.~V. and {Amor{\'\i}n}, Ricardo O. and {Asada}, Yoshihisa and {Atek}, Hakim and {Bezanson}, Rachel and {Bradley}, Larry D. and {Chisholm}, John and {Conselice}, Christopher J. and {Dayal}, Pratika and {Dessauges-Zavadsky}, Miroslava and {Diego}, Jose M. and {Faisst}, Andreas L. and {Fei}, Qinyue and {Frye}, Brenda L. and {Fudamoto}, Yoshinobu and {Furtak}, Lukas J. and {Harikane}, Yuichi and {Hsiao}, Tiger Yu-Yang and {Jim{\'e}nez-Teja}, Yolanda and {Kartaltepe}, Jeyhan S. and {Kiyota}, Tomokazu and {Koekemoer}, Anton M. and {Lagos}, Claudia del P. and {Magdis}, Georgios E. and {Meena}, Ashish Kumar and {Mowla}, Lamiya and {Noirot}, Ga{\"e}l and {Oesch}, Pascal A. and {Ono}, Yoshiaki and {Ortiz}, III, Rafael and {Pan}, Richard and {Papovich}, Casey and {Pierel}, Justin D. and {Ricotti}, Massimo and {Robbins}, Luke and {Schaerer}, Daniel and {Schneider}, Raffaella and {Treu}, Tommaso and {Valentino}, Francesco and {Windhorst}, Rogier A. and {Bauer}, Franz E. and {Bromm}, Volker and {Egami}, Eiichi and {Gonz{\'a}lez-Otero}, Mauro and {Kohno}, Kotaro and {Labbe}, Ivo and {Matthee}, Jorryt and {Mun}, Marcie and {Naidu}, Rohan P. and {Tripodi}, Roberta},
        title = "{VENUS: A Strongly Lensed Clumpy Galaxy at $z\sim11-12$ behind the Galaxy Cluster MACS J0257.1-2325}",
      journal = {arXiv e-prints},
         year = 2025,
        month = nov,
          eid = {arXiv:2511.14483},
        pages = {arXiv:2511.14483},
          doi = {10.48550/arXiv.2511.14483},
archivePrefix = {arXiv},
       eprint = {2511.14483},
 primaryClass = {astro-ph.GA},
       adsurl = {https://ui.adsabs.harvard.edu/abs/2025arXiv251114483N}
}

@ARTICLE{Bradac2025ApJ,
       author = {{Brada{\v{c}}}, Maru{\v{s}}a and {Jude{\v{z}}}, Jon and {Willott}, Chris and {Rihtar{\v{s}}ic}, Gregor and {Martis}, Nicholas S. and {Harshan}, Anishya and {Felicioni}, Giordano and {Asada}, Yoshihisa and {Desprez}, Guillaume and {Clowe}, Douglas and {Gonzalez}, Anthony H. and {Jones}, Christine and {Lemaux}, Brian C. and {Markevitch}, Maxim and {Markov}, Vladan and {Mowla}, Lamiya and {Noirot}, Ga{\"e}l and {Peter}, Annika H.~G. and {Robertson}, Andrew and {Sarrouh}, Ghassan T.~E. and {Sawicki}, Marcin and {Schrabback}, Tim and {Tripodi}, Roberta},
        title = "{Star Formation under a Cosmic Microscope: Highly Magnified z = 11 Galaxy behind the Bullet Cluster}",
      journal = {\apjl},
         year = 2025,
        month = dec,
       volume = {995},
       number = {2},
          eid = {L74},
        pages = {L74},
          doi = {10.3847/2041-8213/ae27d2},
archivePrefix = {arXiv},
       eprint = {2509.20446},
 primaryClass = {astro-ph.GA},
       adsurl = {https://ui.adsabs.harvard.edu/abs/2025ApJ...995L..74B}
}

@ARTICLE{Jakobsen2022,
       author = {{Jakobsen}, P. and {Ferruit}, P. and {Alves de Oliveira}, C. and {Arribas}, S. and {Bagnasco}, G. and {Barho}, R. and {Beck}, T.~L. and {Birkmann}, S. and {B{\"o}ker}, T. and {Bunker}, A.~J. and {Charlot}, S. and {de Jong}, P. and {de Marchi}, G. and {Ehrenwinkler}, R. and {Falcolini}, M. and {Fels}, R. and {Franx}, M. and {Franz}, D. and {Funke}, M. and {Giardino}, G. and {Gnata}, X. and {Holota}, W. and {Honnen}, K. and {Jensen}, P.~L. and {Jentsch}, M. and {Johnson}, T. and {Jollet}, D. and {Karl}, H. and {Kling}, G. and {K{\"o}hler}, J. and {Kolm}, M. -G. and {Kumari}, N. and {Lander}, M.~E. and {Lemke}, R. and {L{\'o}pez-Caniego}, M. and {L{\"u}tzgendorf}, N. and {Maiolino}, R. and {Manjavacas}, E. and {Marston}, A. and {Maschmann}, M. and {Maurer}, R. and {Messerschmidt}, B. and {Moseley}, S.~H. and {Mosner}, P. and {Mott}, D.~B. and {Muzerolle}, J. and {Pirzkal}, N. and {Pittet}, J. -F. and {Plitzke}, A. and {Posselt}, W. and {Rapp}, B. and {Rauscher}, B.~J. and {Rawle}, T. and {Rix}, H. -W. and {R{\"o}del}, A. and {Rumler}, P. and {Sabbi}, E. and {Salvignol}, J. -C. and {Schmid}, T. and {Sirianni}, M. and {Smith}, C. and {Strada}, P. and {te Plate}, M. and {Valenti}, J. and {Wettemann}, T. and {Wiehe}, T. and {Wiesmayer}, M. and {Willott}, C.~J. and {Wright}, R. and {Zeidler}, P. and {Zincke}, C.},
        title = "{The Near-Infrared Spectrograph (NIRSpec) on the James Webb Space Telescope. I. Overview of the instrument and its capabilities}",
      journal = {\aap},
         year = 2022,
        month = may,
       volume = {661},
          eid = {A80},
        pages = {A80},
          doi = {10.1051/0004-6361/202142663},
archivePrefix = {arXiv},
       eprint = {2202.03305},
 primaryClass = {astro-ph.IM},
       adsurl = {https://ui.adsabs.harvard.edu/abs/2022A&A...661A..80J}
}

@ARTICLE{Lacey1993,
       author = {{Lacey}, Cedric and {Cole}, Shaun},
        title = "{Merger rates in hierarchical models of galaxy formation}",
      journal = {\mnras},
         year = 1993,
        month = jun,
       volume = {262},
       number = {3},
        pages = {627-649},
          doi = {10.1093/mnras/262.3.627},
       adsurl = {https://ui.adsabs.harvard.edu/abs/1993MNRAS.262..627L}
}

@ARTICLE{White1991,
       author = {{White}, Simon D.~M. and {Frenk}, Carlos S.},
        title = "{Galaxy Formation through Hierarchical Clustering}",
      journal = {\apj},
         year = 1991,
        month = sep,
       volume = {379},
        pages = {52},
          doi = {10.1086/170483},
       adsurl = {https://ui.adsabs.harvard.edu/abs/1991ApJ...379...52W}
}

@ARTICLE{Patton2002,
       author = {{Patton}, D.~R. and {Pritchet}, C.~J. and {Carlberg}, R.~G. and {Marzke}, R.~O. and {Yee}, H.~K.~C. and {Hall}, P.~B. and {Lin}, H. and {Morris}, S.~L. and {Sawicki}, M. and {Shepherd}, C.~W. and {Wirth}, G.~D.},
        title = "{Dynamically Close Galaxy Pairs and Merger Rate Evolution in the CNOC2 Redshift Survey}",
      journal = {\apj},
         year = 2002,
        month = jan,
       volume = {565},
       number = {1},
        pages = {208-222},
          doi = {10.1086/324543},
archivePrefix = {arXiv},
       eprint = {astro-ph/0109428},
 primaryClass = {astro-ph},
       adsurl = {https://ui.adsabs.harvard.edu/abs/2002ApJ...565..208P}
}

@ARTICLE{Markov2025arXiv,
       author = {{Markov}, V. and {Gallerani}, S. and {Pallottini}, A. and {Bradac}, M. and {Carniani}, S. and {Tripodi}, R. and {Noirot}, G. and {Di Mascia}, F. and {Parlanti}, E. and {Martis}, N.},
        title = "{Unveiling the trends between dust attenuation and galaxy properties at $z \sim 2$-12 with JWST}",
      journal = {arXiv e-prints},
         year = 2025,
        month = apr,
          eid = {arXiv:2504.12378},
        pages = {arXiv:2504.12378},
          doi = {10.48550/arXiv.2504.12378},
archivePrefix = {arXiv},
       eprint = {2504.12378},
 primaryClass = {astro-ph.GA},
       adsurl = {https://ui.adsabs.harvard.edu/abs/2025arXiv250412378M}
}

@software{Brammer2022zndo,
       author = {{Brammer}, Gabe},
        title = "{gbrammer/msaexp: Full working version with 2d drizzling and extraction}",
         year = 2022,
        month = nov,
          eid = {10.5281/zenodo.7299501},
          doi = {10.5281/zenodo.7299501},
      version = {0.3},
    publisher = {Zenodo},
       adsurl = {https://ui.adsabs.harvard.edu/abs/2022zndo...7299501B}
}

@ARTICLE{Song2016ApJ,
       author = {{Song}, Mimi and {Finkelstein}, Steven L. and {Ashby}, Matthew L.~N. and {Grazian}, A. and {Lu}, Yu and {Papovich}, Casey and {Salmon}, Brett and {Somerville}, Rachel S. and {Dickinson}, Mark and {Duncan}, K. and {Faber}, Sandy M. and {Fazio}, Giovanni G. and {Ferguson}, Henry C. and {Fontana}, Adriano and {Guo}, Yicheng and {Hathi}, Nimish and {Lee}, Seong-Kook and {Merlin}, Emiliano and {Willner}, S.~P.},
        title = "{The Evolution of the Galaxy Stellar Mass Function at z = 4-8: A Steepening Low-mass-end Slope with Increasing Redshift}",
      journal = {\apj},
         year = 2016,
        month = jul,
       volume = {825},
       number = {1},
          eid = {5},
        pages = {5},
          doi = {10.3847/0004-637X/825/1/5},
archivePrefix = {arXiv},
       eprint = {1507.05636},
 primaryClass = {astro-ph.GA},
       adsurl = {https://ui.adsabs.harvard.edu/abs/2016ApJ...825....5S}
}

@ARTICLE{DEugenio2024AAP,
       author = {{D'Eugenio}, Francesco and {Maiolino}, Roberto and {Carniani}, Stefano and {Chevallard}, Jacopo and {Curtis-Lake}, Emma and {Witstok}, Joris and {Charlot}, Stephane and {Baker}, William M. and {Arribas}, Santiago and {Boyett}, Kristan and {Bunker}, Andrew J. and {Curti}, Mirko and {Eisenstein}, Daniel J. and {Hainline}, Kevin and {Ji}, Zhiyuan and {Johnson}, Benjamin D. and {Kumari}, Nimisha and {Looser}, Tobias J. and {Nakajima}, Kimihiko and {Nelson}, Erica and {Rieke}, Marcia and {Robertson}, Brant and {Scholtz}, Jan and {Smit}, Renske and {Sun}, Fengwu and {Venturi}, Giacomo and {Tacchella}, Sandro and {{\"U}bler}, Hannah and {Willmer}, Christopher N.~A. and {Willott}, Chris},
        title = "{JADES: Carbon enrichment 350 Myr after the Big Bang}",
      journal = {\aap},
         year = 2024,
        month = sep,
       volume = {689},
          eid = {A152},
        pages = {A152},
          doi = {10.1051/0004-6361/202348636},
archivePrefix = {arXiv},
       eprint = {2311.09908},
 primaryClass = {astro-ph.GA},
       adsurl = {https://ui.adsabs.harvard.edu/abs/2024A&A...689A.152D}
}

@software{Brammer2019ascl,
       author = {{Brammer}, Gabe},
        title = "{Grizli: Grism redshift and line analysis software}",
 howpublished = {Astrophysics Source Code Library, record ascl:1905.001},
         year = 2019,
        month = may,
          eid = {ascl:1905.001},
       adsurl = {https://ui.adsabs.harvard.edu/abs/2019ascl.soft05001B}
}

@ARTICLE{Rihtarsic2025,
       author = {{Rihtar{\v{s}}i{\v{c}}}, G. and {Brada{\v{c}}}, M. and {Desprez}, G. and {Harshan}, A. and {Noirot}, G. and {Estrada-Carpenter}, V. and {Martis}, N.~S. and {Abraham}, R.~G. and {Asada}, Y. and {Brammer}, G. and {Iyer}, K.~G. and {Matharu}, J. and {Mowla}, L. and {Muzzin}, A. and {Sarrouh}, G.~T.~E. and {Sawicki}, M. and {Strait}, V. and {Willott}, C.~J. and {Gledhill}, R. and {Markov}, V. and {Tripodi}, R.},
        title = "{CANUCS: Constraining the MACS J0416.1-2403 strong lensing model with JWST NIRISS, NIRSpec, and NIRCam}",
      journal = {\aap},
         year = 2025,
        month = apr,
       volume = {696},
          eid = {A15},
        pages = {A15},
          doi = {10.1051/0004-6361/202451117},
archivePrefix = {arXiv},
       eprint = {2406.10332},
 primaryClass = {astro-ph.CO},
       adsurl = {https://ui.adsabs.harvard.edu/abs/2025A&A...696A..15R}
}

@ARTICLE{Gledhill2024,
       author = {{Gledhill}, Rachel and {Strait}, Victoria and {Desprez}, Guillaume and {Rihtar{\v{s}}i{\v{c}}}, Gregor and {Brada{\v{c}}}, Maru{\v{s}}a and {Brammer}, Gabriel and {Willott}, Chris J. and {Martis}, Nicholas and {Sawicki}, Marcin and {Noirot}, Ga{\"e}l and {Sarrouh}, Ghassan T.~E. and {Muzzin}, Adam},
        title = "{CANUCS: An Updated Mass and Magnification Model of A370 with JWST}",
      journal = {\apj},
         year = 2024,
        month = oct,
       volume = {973},
       number = {2},
          eid = {77},
        pages = {77},
          doi = {10.3847/1538-4357/ad684a},
archivePrefix = {arXiv},
       eprint = {2403.07062},
 primaryClass = {astro-ph.GA},
       adsurl = {https://ui.adsabs.harvard.edu/abs/2024ApJ...973...77G}
}

@ARTICLE{Sarrouh2026,
       author = {{Sarrouh}, Ghassan T.~E. and {Asada}, Yoshihisa and {Martis}, Nicholas S. and {Willott}, Chris J. and {Iyer}, Kartheik G. and {Noirot}, Ga{\"e}l and {Muzzin}, Adam and {Sawicki}, Marcin and {Brammer}, Gabriel and {Desprez}, Guillaume and {Rihtar{\v{s}}i{\v{c}}}, Gregor and {Zabl}, Johannes and {Abraham}, Roberto and {Brada{\v{c}}}, Maru{\v{s}}a and {Doyon}, Ren{\'e} and {Antwi-Danso}, Jacqueline and {Berek}, Samantha and {Brown}, Westley and {Estrada-Carpenter}, Vince and {Favaro}, Jeremy and {Felicioni}, Giordano and {Forrest}, Ben and {Gaspar}, Gaia and {Gould}, Katriona M.~L. and {Gledhill}, Rachel and {Harshan}, Anishya and {Jahan}, Nusrath and {Jagga}, Naadiyah and {Jude{\v{z}}}, Jon and {Marchesini}, Danilo and {Markov}, Vladan and {Matharu}, Jasleen and {MacFarland}, Shannon and {Merchant}, Maya and {M{\'e}rida}, Rosa M. and {Mowla}, Lamiya and {Myers}, Katherine and {Omori}, Kiyoaki C. and {Pacifici}, Camilla and {Ravindranath}, Swara and {Robbins}, Luke and {Strait}, Victoria and {Sok}, Visal and {Tan}, Vivian Yun Yan and {Tripodi}, Roberta and {Wilson}, Gillian and {Withers}, Sunna},
        title = "{CANUCS/Technicolor Data Release 1: Imaging, Photometry, Slit Spectroscopy, and Stellar Population Parameters}",
      journal = {\apjs},
         year = 2026,
        month = jan,
       volume = {282},
       number = {1},
          eid = {3},
        pages = {3},
          doi = {10.3847/1538-4365/ae1611},
archivePrefix = {arXiv},
       eprint = {2506.21685},
 primaryClass = {astro-ph.GA},
       adsurl = {https://ui.adsabs.harvard.edu/abs/2026ApJS..282....3S}
}

@ARTICLE{Mahler2019,
       author = {{Mahler}, Guillaume and {Sharon}, Keren and {Fox}, Carter and {Coe}, Dan and {Jauzac}, Mathilde and {Strait}, Victoria and {Edge}, Alastair and {Acebron}, Ana and {Andrade-Santos}, Felipe and {Avila}, Roberto J. and {Brada{\v{c}}}, Maru{\v{s}}a and {Bradley}, Larry D. and {Carrasco}, Daniela and {Cerny}, Catherine and {Cibirka}, Nath{\'a}lia and {Czakon}, Nicole G. and {Dawson}, William A. and {Frye}, Brenda L. and {Hoag}, Austin T. and {Huang}, Kuang-Han and {Johnson}, Traci L. and {Jones}, Christine and {Kikuchihara}, Shotaro and {Lam}, Daniel and {Livermore}, Rachael and {Lovisari}, Lorenzo and {Mainali}, Ramesh and {Ogaz}, Sara and {Ouchi}, Masami and {Paterno-Mahler}, Rachel and {Roederer}, Ian U. and {Ryan}, Russell E. and {Salmon}, Brett and {Sendra-Server}, Irene and {Stark}, Daniel P. and {Toft}, Sune and {Trenti}, Michele and {Umetsu}, Keiichi and {Vulcani}, Benedetta and {Zitrin}, Adi},
        title = "{RELICS: Strong Lensing Analysis of MACS J0417.5-1154 and Predictions for Observing the Magnified High-redshift Universe with JWST}",
      journal = {\apj},
         year = 2019,
        month = mar,
       volume = {873},
       number = {1},
          eid = {96},
        pages = {96},
          doi = {10.3847/1538-4357/ab042b},
archivePrefix = {arXiv},
       eprint = {1810.13439},
 primaryClass = {astro-ph.GA},
       adsurl = {https://ui.adsabs.harvard.edu/abs/2019ApJ...873...96M}
}

@ARTICLE{Jauzac2019,
       author = {{Jauzac}, Mathilde and {Mahler}, Guillaume and {Edge}, Alastair C. and {Sharon}, Keren and {Gillman}, Steven and {Ebeling}, Harald and {Harvey}, David and {Richard}, Johan and {Hamer}, Stephen L. and {Fumagalli}, Michele and {Mark Swinbank}, A. and {Kneib}, Jean-Paul and {Massey}, Richard and {Salom{\'e}}, Philippe},
        title = "{The core of the massive cluster merger MACS J0417.5-1154 as seen by VLT/MUSE}",
      journal = {\mnras},
         year = 2019,
        month = mar,
       volume = {483},
       number = {3},
        pages = {3082-3097},
          doi = {10.1093/mnras/sty3312},
archivePrefix = {arXiv},
       eprint = {1811.02505},
 primaryClass = {astro-ph.GA},
       adsurl = {https://ui.adsabs.harvard.edu/abs/2019MNRAS.483.3082J}
}

@ARTICLE{Eliasdottir2007arXiv,
       author = {{El{\'\i}asd{\'o}ttir}, {\'A}rd{\'\i}s and {Limousin}, Marceau and {Richard}, Johan and {Hjorth}, Jens and {Kneib}, Jean-Paul and {Natarajan}, Priya and {Pedersen}, Kristian and {Jullo}, Eric and {Paraficz}, Danuta},
        title = "{Where is the matter in the Merging Cluster Abell 2218?}",
      journal = {arXiv e-prints},
         year = 2007,
        month = oct,
          eid = {arXiv:0710.5636},
        pages = {arXiv:0710.5636},
          doi = {10.48550/arXiv.0710.5636},
archivePrefix = {arXiv},
       eprint = {0710.5636},
 primaryClass = {astro-ph},
       adsurl = {https://ui.adsabs.harvard.edu/abs/2007arXiv0710.5636E}
}

@ARTICLE{Jullo2007,
   author = {{Jullo}, E. and {Kneib}, J.-P. and {Limousin}, M. and {El{\'{\i}}asd{\'o}ttir}, {\'A}. and 
	{Marshall}, P.~J. and {Verdugo}, T.},
    title = "{A Bayesian approach to strong lensing modelling of galaxy clusters}",
  journal = {New Journal of Physics},
archivePrefix = "arXiv",
   eprint = {0706.0048},
     year = 2007,
    month = dec,
   volume = 9,
    pages = {447},
      doi = {10.1088/1367-2630/9/12/447},
   adsurl = {http://adsabs.harvard.edu/abs/2007NJPh....9..447J}
}

@PHDTHESIS{Kneib1993,
       author = {{Kneib}, J. -P.},
       school = {-},
         year = 1993,
        month = jan,
       adsurl = {https://ui.adsabs.harvard.edu/abs/1993PhDT.......189K}
}

@ARTICLE{Estrada-carpenter2024,
       author = {{Estrada-Carpenter}, Vicente and {Sawicki}, Marcin and {Brammer}, Gabe and {Desprez}, Guillaume and {Abraham}, Roberto and {Asada}, Yoshihisa and {Brada{\v{c}}}, Maru{\v{s}}a and {Iyer}, Kartheik G. and {Martis}, Nicholas S. and {Matharu}, Jasleen and {Mowla}, Lamiya and {Muzzin}, Adam and {Noirot}, Ga{\"e}l and {Sarrouh}, Ghassan T.~E. and {Strait}, Victoria and {Willott}, Chris J.},
        title = "{When, where, and how star formation happens in a galaxy pair at cosmic noon using CANUCS JWST/NIRISS grism spectroscopy}",
      journal = {\mnras},
         year = 2024,
        month = jul,
       volume = {532},
       number = {1},
        pages = {577-591},
          doi = {10.1093/mnras/stae1368},
archivePrefix = {arXiv},
       eprint = {2406.15551},
 primaryClass = {astro-ph.GA},
       adsurl = {https://ui.adsabs.harvard.edu/abs/2024MNRAS.532..577E}
}

@ARTICLE{Weaver2023AAS,
       author = {{Weaver}, J.~R. and {Davidzon}, I. and {Toft}, S. and {Ilbert}, O. and {McCracken}, H.~J. and {Gould}, K.~M.~L. and {Jespersen}, C.~K. and {Steinhardt}, C. and {Lagos}, C.~D.~P. and {Capak}, P.~L. and {Casey}, C.~M. and {Chartab}, N. and {Faisst}, A.~L. and {Hayward}, C.~C. and {Kartaltepe}, J.~S. and {Kauffmann}, O.~B. and {Koekemoer}, A.~M. and {Kokorev}, V. and {Laigle}, C. and {Liu}, D. and {Long}, A. and {Magdis}, G.~E. and {McPartland}, C.~J.~R. and {Milvang-Jensen}, B. and {Mobasher}, B. and {Moneti}, A. and {Peng}, Y. and {Sanders}, D.~B. and {Shuntov}, M. and {Sneppen}, A. and {Valentino}, F. and {Zalesky}, L. and {Zamorani}, G.},
        title = "{COSMOS2020: The galaxy stellar mass function. The assembly and star formation cessation of galaxies at 0.2< z {\ensuremath{\leq}} 7.5}",
      journal = {\aap},
         year = 2023,
        month = sep,
       volume = {677},
          eid = {A184},
        pages = {A184},
          doi = {10.1051/0004-6361/202245581},
archivePrefix = {arXiv},
       eprint = {2212.02512},
 primaryClass = {astro-ph.GA},
       adsurl = {https://ui.adsabs.harvard.edu/abs/2023A&A...677A.184W}
}

@ARTICLE{Baldry2012MNRAS,
       author = {{Baldry}, I.~K. and {Driver}, S.~P. and {Loveday}, J. and {Taylor}, E.~N. and {Kelvin}, L.~S. and {Liske}, J. and {Norberg}, P. and {Robotham}, A.~S.~G. and {Brough}, S. and {Hopkins}, A.~M. and {Bamford}, S.~P. and {Peacock}, J.~A. and {Bland-Hawthorn}, J. and {Conselice}, C.~J. and {Croom}, S.~M. and {Jones}, D.~H. and {Parkinson}, H.~R. and {Popescu}, C.~C. and {Prescott}, M. and {Sharp}, R.~G. and {Tuffs}, R.~J.},
        title = "{Galaxy And Mass Assembly (GAMA): the galaxy stellar mass function at z < 0.06}",
      journal = {\mnras},
         year = 2012,
        month = mar,
       volume = {421},
       number = {1},
        pages = {621-634},
          doi = {10.1111/j.1365-2966.2012.20340.x},
archivePrefix = {arXiv},
       eprint = {1111.5707},
 primaryClass = {astro-ph.CO},
       adsurl = {https://ui.adsabs.harvard.edu/abs/2012MNRAS.421..621B}
}

@ARTICLE{Tan2024arXiv,
       author = {{Tan}, Vivian Yun Yan and {Muzzin}, Adam and {Sarrouh}, Ghassan T.~E. and {Antwi-Danso}, Jacqueline and {Sok}, Visal and {Jagga}, Naadiyah and {Rihtar{\v{s}}i{\v{c}}}, Gregor and {Brown}, Westley and {Abraham}, Roberto and {Asada}, Yoshihisa and {Desprez}, Guillaume and {Iyer}, Kartheik and {Martis}, Nicholas S. and {M{\'e}rida}, Rosa M. and {Mowla}, Lamiya A. and {Noirot}, Ga{\"e}l and {Omori}, Kiyoaki Christopher and {Sawicki}, Marcin and {Tripodi}, Roberta and {Willott}, Chris J.},
        title = "{Resolved Mass Assembly and Star Formation in Milky Way Progenitors since z = 5 from JWST/CANUCS: From Clumps and Mergers to Well-ordered Disks}",
      journal = {\apj},
         year = 2025,
        month = nov,
       volume = {994},
       number = {1},
          eid = {94},
        pages = {94},
          doi = {10.3847/1538-4357/ae0ffe},
archivePrefix = {arXiv},
       eprint = {2412.07829},
 primaryClass = {astro-ph.GA},
       adsurl = {https://ui.adsabs.harvard.edu/abs/2025ApJ...994...94T}
}

@ARTICLE{Eldridge2017,
       author = {{Eldridge}, J.~J. and {Stanway}, E.~R. and {Xiao}, L. and {McClelland}, L.~A.~S. and {Taylor}, G. and {Ng}, M. and {Greis}, S.~M.~L. and {Bray}, J.~C.},
        title = "{Binary Population and Spectral Synthesis Version 2.1: Construction, Observational Verification, and New Results}",
      journal = {\pasa},
         year = 2017,
        month = nov,
       volume = {34},
          eid = {e058},
        pages = {e058},
          doi = {10.1017/pasa.2017.51},
archivePrefix = {arXiv},
       eprint = {1710.02154},
 primaryClass = {astro-ph.SR},
       adsurl = {https://ui.adsabs.harvard.edu/abs/2017PASA...34...58E}
}

@ARTICLE{Leja2019,
       author = {{Leja}, Joel and {Carnall}, Adam C. and {Johnson}, Benjamin D. and {Conroy}, Charlie and {Speagle}, Joshua S.},
        title = "{How to Measure Galaxy Star Formation Histories. II. Nonparametric Models}",
      journal = {\apj},
         year = 2019,
        month = may,
       volume = {876},
       number = {1},
          eid = {3},
        pages = {3},
          doi = {10.3847/1538-4357/ab133c},
archivePrefix = {arXiv},
       eprint = {1811.03637},
 primaryClass = {astro-ph.GA},
       adsurl = {https://ui.adsabs.harvard.edu/abs/2019ApJ...876....3L}
}

@ARTICLE{Berg2021,
       author = {{Berg}, Danielle A. and {Chisholm}, John and {Erb}, Dawn K. and {Skillman}, Evan D. and {Pogge}, Richard W. and {Olivier}, Grace M.},
        title = "{Characterizing Extreme Emission-line Galaxies. I. A Four-zone Ionization Model for Very High-ionization Emission}",
      journal = {\apj},
         year = 2021,
        month = dec,
       volume = {922},
       number = {2},
          eid = {170},
        pages = {170},
          doi = {10.3847/1538-4357/ac141b},
archivePrefix = {arXiv},
       eprint = {2105.12765},
 primaryClass = {astro-ph.GA},
       adsurl = {https://ui.adsabs.harvard.edu/abs/2021ApJ...922..170B}
}

@ARTICLE{Campbell1986,
       author = {{Campbell}, Alison and {Terlevich}, Roberto and {Melnick}, Jorge},
        title = "{The stellar populations and evolution of H II galaxies - I. High signal-to-noise optical spectroscopy.}",
      journal = {\mnras},
         year = 1986,
        month = dec,
       volume = {223},
        pages = {811-825},
          doi = {10.1093/mnras/223.4.811},
       adsurl = {https://ui.adsabs.harvard.edu/abs/1986MNRAS.223..811C}
}

@ARTICLE{Curti_2023MNRAS,
       author = {{Curti}, Mirko and {D'Eugenio}, Francesco and {Carniani}, Stefano and {Maiolino}, Roberto and {Sandles}, Lester and {Witstok}, Joris and {Baker}, William M. and {Bennett}, Jake S. and {Piotrowska}, Joanna M. and {Tacchella}, Sandro and {Charlot}, Stephane and {Nakajima}, Kimihiko and {Maheson}, Gabriel and {Mannucci}, Filippo and {Amiri}, Amirnezam and {Arribas}, Santiago and {Belfiore}, Francesco and {Bonaventura}, Nina R. and {Bunker}, Andrew J. and {Chevallard}, Jacopo and {Cresci}, Giovanni and {Curtis-Lake}, Emma and {Hayden-Pawson}, Connor and {Jones}, Gareth C. and {Kumari}, Nimisha and {Laseter}, Isaac and {Looser}, Tobias J. and {Marconi}, Alessandro and {Maseda}, Michael V. and {Scholtz}, Jan and {Smit}, Renske and {{\"U}bler}, Hannah and {Wallace}, Imaan E.~B.},
        title = "{The chemical enrichment in the early Universe as probed by JWST via direct metallicity measurements at z {\ensuremath{\sim}} 8}",
      journal = {\mnras},
         year = 2023,
        month = jan,
       volume = {518},
       number = {1},
        pages = {425-438},
          doi = {10.1093/mnras/stac2737},
archivePrefix = {arXiv},
       eprint = {2207.12375},
 primaryClass = {astro-ph.GA},
       adsurl = {https://ui.adsabs.harvard.edu/abs/2023MNRAS.518..425C}
}

@ARTICLE{Sanders_2024ApJ,
       author = {{Sanders}, Ryan L. and {Shapley}, Alice E. and {Topping}, Michael W. and {Reddy}, Naveen A. and {Brammer}, Gabriel B.},
        title = "{Direct T $_{e}$-based Metallicities of z = 2{\textendash}9 Galaxies with JWST/NIRSpec: Empirical Metallicity Calibrations Applicable from Reionization to Cosmic Noon}",
      journal = {\apj},
         year = 2024,
        month = feb,
       volume = {962},
       number = {1},
          eid = {24},
        pages = {24},
          doi = {10.3847/1538-4357/ad15fc},
archivePrefix = {arXiv},
       eprint = {2303.08149},
 primaryClass = {astro-ph.GA},
       adsurl = {https://ui.adsabs.harvard.edu/abs/2024ApJ...962...24S}
}

@ARTICLE{Izotov2006,
       author = {{Izotov}, Y.~I. and {Stasi{\'n}ska}, G. and {Meynet}, G. and {Guseva}, N.~G. and {Thuan}, T.~X.},
        title = "{The chemical composition of metal-poor emission-line galaxies in the Data Release 3 of the Sloan Digital Sky Survey}",
      journal = {\aap},
         year = 2006,
        month = mar,
       volume = {448},
       number = {3},
        pages = {955-970},
          doi = {10.1051/0004-6361:20053763},
archivePrefix = {arXiv},
       eprint = {astro-ph/0511644},
 primaryClass = {astro-ph},
       adsurl = {https://ui.adsabs.harvard.edu/abs/2006A&A...448..955I}
}

@ARTICLE{Isobe2023,
       author = {{Isobe}, Yuki and {Ouchi}, Masami and {Nakajima}, Kimihiko and {Harikane}, Yuichi and {Ono}, Yoshiaki and {Xu}, Yi and {Zhang}, Yechi and {Umeda}, Hiroya},
        title = "{Redshift Evolution of Electron Density in the Interstellar Medium at z   0-9 Uncovered with JWST/NIRSpec Spectra and Line-spread Function Determinations}",
      journal = {\apj},
         year = 2023,
        month = oct,
       volume = {956},
       number = {2},
          eid = {139},
        pages = {139},
          doi = {10.3847/1538-4357/acf376},
archivePrefix = {arXiv},
       eprint = {2301.06811},
 primaryClass = {astro-ph.GA},
       adsurl = {https://ui.adsabs.harvard.edu/abs/2023ApJ...956..139I}
}

@article{Calzetti2000, title={The Dust Content and Opacity of Actively Star-forming Galaxies}, volume={533}, ISSN={0004-637X}, DOI={10.1086/308692}, abstractNote={We present far-infrared (FIR) photometry at 150 and 205 μm of eight low-redshift starburst galaxies obtained with the Infrared Space Observatory (ISO) ISOPHOT. Five of the eight galaxies are detected in both wave bands, and these data are used, in conjunction with IRAS archival photometry, to model the dust emission at λ>~40
μm. The FIR spectral energy distributions (SEDs) are best fitted by a combination of two modified Planck functions, with T~40-55 K (warm dust) and T~20-23 K (cool dust) and with a dust emissivity index ɛ=2.
The cool dust can be a major contributor to the FIR emission of
starburst galaxies, representing up to 60% of the total flux. This component is heated not only by the general interstellar radiation field, but also by the starburst itself. The cool dust mass is up to ~150 times larger than the warm dust mass, bringing the gas-to-dust ratios of the starbursts in our sample close to Milky Way values, once rescaled for the appropriate metallicity. The ratio between the total dust FIR emission in the range 1-1000 μm and the IRAS FIR emission in the range 40-120 μm is ~1.75, with small variations from galaxy to galaxy. This ratio is about 40% larger than previously inferred from data at millimeter wavelengths. Although the galaxies in our sample are generally classified as ``UV bright,’’ for four of them the UV energy emerging shortward of 0.2 μm is less than 15% of the FIR energy. On average, about 30% of the bolometric flux is coming out in the
UV-to-near-IR wavelength range; the rest is emitted in the FIR. Energy balance calculations show that the FIR emission predicted by the dust reddening of the UV-to-near-IR stellar emission is within a factor of ~2 of the observed value in individual galaxies and within 20% when averaged over a large sample. If our sample of local starbursts is representative of high-redshift (z>~1), UV-bright, star-forming
galaxies, these galaxies’ FIR emission will be generally undetected in submillimeter surveys, unless (1) their bolometric luminosity is comparable to or larger than that of ultraluminous FIR galaxies and (2) their FIR SED contains a cool dust component. Based on observations with ISO, an ESA project with instruments funded by ESA member states (especially the PI countries: France, Germany, the Netherlands, and the United Kingdom) with the participation of ISAS and NASA.}, journal={ApJ}, author={Calzetti, Daniela and Armus, Lee and Bohlin, Ralph C. and Kinney, Anne L. and Koornneef, Jan and Storchi-Bergmann, Thaisa}, year={2000}, month={Apr}, pages={682} }

@BOOK{Osterbrock2006agnagn,
       author = {{Osterbrock}, Donald E. and {Ferland}, Gary J.},
        title = "{Astrophysics of gaseous nebulae and active galactic nuclei}",
         year = 2006,
       adsurl = {https://ui.adsabs.harvard.edu/abs/2006agna.book.....O}
}

@ARTICLE{Horne1986PASP,
       author = {{Horne}, K.},
        title = "{An optimal extraction algorithm for CCD spectroscopy.}",
      journal = {\pasp},
         year = 1986,
        month = jun,
       volume = {98},
        pages = {609-617},
          doi = {10.1086/131801},
       adsurl = {https://ui.adsabs.harvard.edu/abs/1986PASP...98..609H}
}

@ARTICLE{Desprez2024MNRAS,
       author = {{Desprez}, Guillaume and {Martis}, Nicholas S. and {Asada}, Yoshihisa and {Sawicki}, Marcin and {Willott}, Chris J. and {Muzzin}, Adam and {Abraham}, Roberto G. and {Brada{\v{c}}}, Maru{\v{s}}a and {Brammer}, Gabe and {Estrada-Carpenter}, Vicente and {Iyer}, Kartheik G. and {Matharu}, Jasleen and {Mowla}, Lamiya and {Noirot}, Ga{\"e}l and {Sarrouh}, Ghassan T.~E. and {Strait}, Victoria and {Gledhill}, Rachel and {Rihtar{\v{s}}i{\v{c}}}, Gregor},
        title = "{{\ensuremath{\Lambda}}CDM not dead yet: massive high-z Balmer break galaxies are less common than previously reported}",
      journal = {\mnras},
         year = 2024,
        month = may,
       volume = {530},
       number = {3},
        pages = {2935-2952},
          doi = {10.1093/mnras/stae1084},
archivePrefix = {arXiv},
       eprint = {2310.03063},
 primaryClass = {astro-ph.GA},
       adsurl = {https://ui.adsabs.harvard.edu/abs/2024MNRAS.530.2935D}
}

@ARTICLE{Ferruit2022,
       author = {{Ferruit}, P. and {Jakobsen}, P. and {Giardino}, G. and {Rawle}, T. and {Alves de Oliveira}, C. and {Arribas}, S. and {Beck}, T.~L. and {Birkmann}, S. and {B{\"o}ker}, T. and {Bunker}, A.~J. and {Charlot}, S. and {de Marchi}, G. and {Franx}, M. and {Henry}, A. and {Karakla}, D. and {Kassin}, S.~A. and {Kumari}, N. and {L{\'o}pez-Caniego}, M. and {L{\"u}tzgendorf}, N. and {Maiolino}, R. and {Manjavacas}, E. and {Marston}, A. and {Moseley}, S.~H. and {Muzerolle}, J. and {Pirzkal}, N. and {Rauscher}, B. and {Rix}, H. -W. and {Sabbi}, E. and {Sirianni}, M. and {te Plate}, M. and {Valenti}, J. and {Willott}, C.~J. and {Zeidler}, P.},
        title = "{The Near-Infrared Spectrograph (NIRSpec) on the James Webb Space Telescope. II. Multi-object spectroscopy (MOS)}",
      journal = {\aap},
         year = 2022,
        month = may,
       volume = {661},
          eid = {A81},
        pages = {A81},
          doi = {10.1051/0004-6361/202142673},
archivePrefix = {arXiv},
       eprint = {2202.03306},
 primaryClass = {astro-ph.IM},
       adsurl = {https://ui.adsabs.harvard.edu/abs/2022A&A...661A..81F}
}

@ARTICLE{Martis2024ApJ,
       author = {{Martis}, Nicholas S. and {Sarrouh}, Ghassan T.~E. and {Willott}, Chris J. and {Abraham}, Roberto and {Asada}, Yoshihisa and {Brada{\v{c}}}, Maru{\v{s}}a and {Brammer}, Gabriel B. and {Desprez}, Guillaume and {Harshan}, Anishya and {Muzzin}, Adam and {Noirot}, Ga{\"e}l and {Rihtar{\v{s}}i{\v{c}}}, Gregor and {Sawicki}, Marcin and {Strait}, Victoria},
        title = "{Modeling and Subtracting Diffuse Cluster Light in JWST Images: A Relation between the Spatial Distribution of Globular Clusters, Dwarf Galaxies, and Intracluster Light in the Lensing Cluster SMACS 0723}",
      journal = {\apj},
         year = 2024,
        month = nov,
       volume = {975},
       number = {1},
          eid = {76},
        pages = {76},
          doi = {10.3847/1538-4357/ad7735},
archivePrefix = {arXiv},
       eprint = {2401.01945},
 primaryClass = {astro-ph.GA},
       adsurl = {https://ui.adsabs.harvard.edu/abs/2024ApJ...975...76M}
}

@ARTICLE{Adamo2024Natur,
       author = {{Adamo}, Angela and {Bradley}, Larry D. and {Vanzella}, Eros and {Claeyssens}, Ad{\'e}la{\"\i}de and {Welch}, Brian and {Diego}, Jose M. and {Mahler}, Guillaume and {Oguri}, Masamune and {Sharon}, Keren and {Abdurro'uf} and {Hsiao}, Tiger Yu-Yang and {Xu}, Xinfeng and {Messa}, Matteo and {Lassen}, Augusto E. and {Zackrisson}, Erik and {Brammer}, Gabriel and {Coe}, Dan and {Kokorev}, Vasily and {Ricotti}, Massimo and {Zitrin}, Adi and {Fujimoto}, Seiji and {Inoue}, Akio K. and {Resseguier}, Tom and {Rigby}, Jane R. and {Jim{\'e}nez-Teja}, Yolanda and {Windhorst}, Rogier A. and {Hashimoto}, Takuya and {Tamura}, Yoichi},
        title = "{Bound star clusters observed in a lensed galaxy 460 Myr after the Big Bang}",
      journal = {\nat},
         year = 2024,
        month = aug,
       volume = {632},
       number = {8025},
        pages = {513-516},
          doi = {10.1038/s41586-024-07703-7},
archivePrefix = {arXiv},
       eprint = {2401.03224},
 primaryClass = {astro-ph.GA},
       adsurl = {https://ui.adsabs.harvard.edu/abs/2024Natur.632..513A}
}

@ARTICLE{Fujimoto2024arxiv,
       author = {{Fujimoto}, S. and {Ouchi}, M. and {Kohno}, K. and {Valentino}, F. and {Gim{\'e}nez-Arteaga}, C. and {Brammer}, G.~B. and {Furtak}, L.~J. and {Kohandel}, M. and {Oguri}, M. and {Pallottini}, A. and {Richard}, J. and {Zitrin}, A. and {Bauer}, F.~E. and {Boylan-Kolchin}, M. and {Dessauges-Zavadsky}, M. and {Egami}, E. and {Finkelstein}, S.~L. and {Ma}, Z. and {Smail}, I. and {Watson}, D. and {Hutchison}, T.~A. and {Rigby}, J.~R. and {Welch}, B.~D. and {Ao}, Y. and {Bradley}, L.~D. and {Caminha}, G.~B. and {Caputi}, K.~I. and {Espada}, D. and {Endsley}, R. and {Fudamoto}, Y. and {Gonz{\'a}lez-L{\'o}pez}, J. and {Hatsukade}, B. and {Koekemoer}, A.~M. and {Kokorev}, V. and {Laporte}, N. and {Lee}, M. and {Magdis}, G.~E. and {Ono}, Y. and {Rizzo}, F. and {Shibuya}, T. and {Shimasaku}, K. and {Sun}, F. and {Toft}, S. and {Umehata}, H. and {Wang}, T. and {Yajima}, H.},
        title = "{Primordial rotating disk composed of at least 15 dense star-forming clumps at cosmic dawn}",
      journal = {Nature Astronomy},
         year = 2025,
        month = oct,
       volume = {9},
        pages = {1553-1567},
          doi = {10.1038/s41550-025-02592-w},
archivePrefix = {arXiv},
       eprint = {2402.18543},
 primaryClass = {astro-ph.GA},
       adsurl = {https://ui.adsabs.harvard.edu/abs/2025NatAs...9.1553F}
}

@ARTICLE{Behroozi2013ApJ,
       author = {{Behroozi}, Peter S. and {Marchesini}, Danilo and {Wechsler}, Risa H. and {Muzzin}, Adam and {Papovich}, Casey and {Stefanon}, Mauro},
        title = "{Using Cumulative Number Densities to Compare Galaxies across Cosmic Time}",
      journal = {\apjl},
         year = 2013,
        month = nov,
       volume = {777},
       number = {1},
          eid = {L10},
        pages = {L10},
          doi = {10.1088/2041-8205/777/1/L10},
archivePrefix = {arXiv},
       eprint = {1308.3232},
 primaryClass = {astro-ph.CO},
       adsurl = {https://ui.adsabs.harvard.edu/abs/2013ApJ...777L..10B}
}

@ARTICLE{Witten2024NatAs,
       author = {{Witten}, Callum and {Laporte}, Nicolas and {Martin-Alvarez}, Sergio and {Sijacki}, Debora and {Yuan}, Yuxuan and {Haehnelt}, Martin G. and {Baker}, William M. and {Dunlop}, James S. and {Ellis}, Richard S. and {Grogin}, Norman A. and {Illingworth}, Garth and {Katz}, Harley and {Koekemoer}, Anton M. and {Magee}, Daniel and {Maiolino}, Roberto and {McClymont}, William and {P{\'e}rez-Gonz{\'a}lez}, Pablo G. and {Pusk{\'a}s}, D{\'a}vid and {Roberts-Borsani}, Guido and {Santini}, Paola and {Simmonds}, Charlotte},
        title = "{Deciphering Lyman-{\ensuremath{\alpha}} emission deep into the epoch of reionization}",
      journal = {Nature Astronomy},
         year = 2024,
        month = mar,
       volume = {8},
        pages = {384-396},
          doi = {10.1038/s41550-023-02179-3},
archivePrefix = {arXiv},
       eprint = {2303.16225},
 primaryClass = {astro-ph.GA},
       adsurl = {https://ui.adsabs.harvard.edu/abs/2024NatAs...8..384W}
}

@ARTICLE{Dome2024MNRAS,
       author = {{Dome}, Tibor and {Tacchella}, Sandro and {Fialkov}, Anastasia and {Ceverino}, Daniel and {Dekel}, Avishai and {Ginzburg}, Omri and {Lapiner}, Sharon and {Looser}, Tobias J.},
        title = "{Mini-quenching of z = 4-8 galaxies by bursty star formation}",
      journal = {\mnras},
         year = 2024,
        month = jan,
       volume = {527},
       number = {2},
        pages = {2139-2151},
          doi = {10.1093/mnras/stad3239},
archivePrefix = {arXiv},
       eprint = {2305.07066},
 primaryClass = {astro-ph.GA},
       adsurl = {https://ui.adsabs.harvard.edu/abs/2024MNRAS.527.2139D}
}

@ARTICLE{Roberts-Borsani2024ApJ,
       author = {{Roberts-Borsani}, Guido and {Treu}, Tommaso and {Shapley}, Alice and {Fontana}, Adriano and {Pentericci}, Laura and {Castellano}, Marco and {Morishita}, Takahiro and {Bergamini}, Pietro and {Rosati}, Piero},
        title = "{Between the Extremes: A JWST Spectroscopic Benchmark for High-redshift Galaxies Using {\ensuremath{\sim}}500 Confirmed Sources at z {\ensuremath{\geq}} 5}",
      journal = {\apj},
         year = 2024,
        month = dec,
       volume = {976},
       number = {2},
          eid = {193},
        pages = {193},
          doi = {10.3847/1538-4357/ad85d3},
archivePrefix = {arXiv},
       eprint = {2403.07103},
 primaryClass = {astro-ph.GA},
       adsurl = {https://ui.adsabs.harvard.edu/abs/2024ApJ...976..193R}
}

@ARTICLE{Cameron2024MNRAS,
       author = {{Cameron}, Alex J. and {Katz}, Harley and {Witten}, Callum and {Saxena}, Aayush and {Laporte}, Nicolas and {Bunker}, Andrew J.},
        title = "{Nebular dominated galaxies: insights into the stellar initial mass function at high redshift}",
      journal = {\mnras},
         year = 2024,
        month = oct,
       volume = {534},
       number = {1},
        pages = {523-543},
          doi = {10.1093/mnras/stae1547},
archivePrefix = {arXiv},
       eprint = {2311.02051},
 primaryClass = {astro-ph.GA},
       adsurl = {https://ui.adsabs.harvard.edu/abs/2024MNRAS.534..523C}
}

@ARTICLE{Chabrier2003,
       author = {{Chabrier}, Gilles},
        title = "{Galactic Stellar and Substellar Initial Mass Function}",
      journal = {\pasp},
         year = 2003,
        month = jul,
       volume = {115},
       number = {809},
        pages = {763-795},
          doi = {10.1086/376392},
archivePrefix = {arXiv},
       eprint = {astro-ph/0304382},
 primaryClass = {astro-ph},
       adsurl = {https://ui.adsabs.harvard.edu/abs/2003PASP..115..763C}
}
\bibliographystyle{aasjournalv7}



\end{document}